\documentclass{iau} 
\usepackage{graphicx}
\usepackage{txfonts}
\usepackage{natbib}
\usepackage{longtable} 

\newcommand\Earth{\hbox{$\oplus$}}

\usepackage[T1]{fontenc}
\usepackage{ae,aecompl}

\title[INPOP20a and Bepi-Colombo] 
{Evolution of INPOP planetary ephemerides and Bepi-Colombo simulations}

\author[A. Fienga \& al.]   
{A. Fienga$^{1,2}$
\and L. Bigot $^{4}$
\and D. Mary $^{4}$
\and P. Deram$^{1}$
\and A. Di Ruscio$^{3,1}$
\and L. Bernus $^{2,1}$
\and M. Gastineau$^{2}$
\and J. Laskar$^{2}$
}

\affiliation{$^1$G\'eoAzur, Observatoire C\^ote d'Azur, Universit\'e C\^ote d'Azur, CNRS, 250 Av. A. Einstein, Valbonne, 06560, France \\ email: {\tt agnes.fienga@oca.eu} \\[\affilskip]
$^2$IMCCE, Observatoire de Paris, PSL University, CNRS, Sorbonne Universit\'e, 77 avenue Denfert-Rochereau, Paris, 75014, France \\
$^3$Dipartimento di Ingegneria Meccanica e Aerospaziale, Sapienza Universit\`a di Roma, via Eudossiana 18, 00184 Rome, Italy \\
$^4$Lagrange, Universit\'e C\^ote d'Azur, Observatoire de la C\^ote d'Azur, CNRS, Lagrange UMR 7293, CS 34229,  06304, Nice Cedex 4, France

}

\pubyear{2015}
\volume{xxx}  
\jname{Title of your IAU Symposium}
\editors{A.C. Editor, B.D. Editor \& C.E. Editor, eds.}
\begin{document}

\maketitle

\begin{abstract}
We give here a detailed description of the latest INPOP planetary ephemerides INPOP20a.  We test the sensitivity of the Sun oblateness determination obtained with INPOP to different models for the Sun core rotation. We also present new evaluations of possible GRT violations with the PPN parameters $\beta$,$\gamma$ and  $\dot{\mu}/\mu$. With a new method for selecting acceptable alternative ephemerides we provide conservative limits of about $7.16 \times 10^{-5}$ and  $7.49 \times 10^{-5}$ for $\beta-1$ and $\gamma-1$ respectively using the present day planetary data samples.  We also present simulations of Bepi-Colombo range tracking data and their impact on planetary ephemeris construction. We show that the use of future BC range observations should improve these estimates, in particular $\gamma$. Finally, interesting perspectives for the detection of the Sun core rotation seem to be reachable thanks to the BC mission and its accurate range measurements in the GRT frame.
\end{abstract}

\firstsection 

\section{Introduction}

The INPOP (Int\'egrateur Num\'erique Plan\'etaire de l'Observatoire de Paris) planetary ephemeris has started to be built in 2003.  
It consists in numerically integrating the Einstein-Imfeld-Hoffman equations of motion and relativistic time-scale definitions (TT,TDB,TCG,TCB) for the eight planets of our solar system, and the Moon (orbit and rotation) \citep{2009A&A...507.1675F}. In our latest version, INPOP20a presented in this paper, orbits of about 343 Main Belt asteroids and 500 Trans-Neptunian Objects (TNO) are also integrated. An adjustment of 402 parameters, including planetary initial conditions, gravitational mass of the Sun and its oblateness, the Earth Moon mass ratio, 343 Main Belt asteroid masses and one global mass for the 500 TNOs, obtained in using 150,000 observations from spacecraft tracking and ground-based optical observations.


In Solar physics, important questions are still pending. In particular, the solar differential rotation has been determined by helioseismology with great precision using the acoustic modes (p-modes) trapped into the solar cavity \citep[for a review]{2003ARA&A..41..599T}.  It was shown that rotation depends on both the depth $r$ and the latitude $\theta$ in the convection zone ($r > 0.71  R_{\odot}$). In the radiative zone, the rotation is near solid (uniform) with a rotation rate of about $\Omega_0/2\pi = 435\,  {\rm nHz}$ \citep{2003ApJ...586..650K}. The rotation of the solar core in still a matter of debate since it is hardly inferred by p-modes. Indeed, since the sensitivity of the p-modes decreases towards the core due to their decreasing mode amplitude at large sound speed, helioseismic inversions were done with precision, from the surface down to about $r\sim 0.3\,R_\odot $. Several instruments both from ground (networks of telescopes around the world) and space missions have tried to infer the solar core rotation using p-modes, but they diverge below $r/R_\odot  \sim 0.25$ \citep[see][for a comparison of rotation profiles]{2003LNP...599...31D}. While space measurements from GOLF on board on the SOHO spacecraft \citep{1998ESASP.418..329R, 2004SoPh..220..269G} and ground-based observations from the networks IRIS \citep{1996SoPh..166....1L} and GONG \citep{1998ESASP.418..193G}  all show an increase of the rotation in the core, the Bison network result is in favor of a decrease \citep{1999MNRAS.308..405C, 2015SSRv..196...15G}.  
 It is worth mentioning that tentative detections of gravity mode (g-modes)  using the GOLF data have also led to a very fast core rotation roughly 3-4 times the value of the radiative zone \citep{2007Sci...316.1591G,2017A&A...604A..40F,2018A&A...612L...1F}. These claimed detections are still questioned by the community and must be taken with care \citep[see][for  discussions about the recent analysis of Fossat et al. 2017]{2018SoPh..293...95S,2019ApJ...877...42S,2019A&A...624A.106A}.  A redetermination of the solar rotation using decades of seismic data up to the most recent ones is under investigation \citep{Christensen-Dalsgaard2021} since the exact rotation profile in the core is important to understand the past history of our Sun, its angular momentum transport, and the role of the magnetic field on solar structure. In view of these uncertainties of the rotation in the core, we have tested the possibility of using INPOP ephemerides  to disentangle these scenarios : faster, slower, or solid rotation in the solar core, through its impact on the Sun gravitational oblateness, $J^{\odot}_{2}$.

For this work, we use the latest update of the INPOP planetary ephemerides, INPOP20a, for obtaining for the Post Parametrized Newtonian (PPN) parameters $\beta$ and $\gamma$ as we did with INPOP15a \citep{2015CeMDA.123..325F} but with a new criterion for ephemeris selection, a more accurate ephemeris and a more complete dynamical modeling. We also give constraints on the Sun oblateness related to different hypotheses of the Sun core rotation, together with new limits for the Sun gravitational mass loss, based on planetary orbits. We also consider the future evolution of INPOP in simulating Mercury-Earth range observations from the Bepi-Colombo mission and in deducing subsequent improvements for the ephemerides and General Relativity Theory (GRT) tests.

Sect. \ref{sec1} introduces the INPOP20a planetary ephemeris, describing the modifications brought in the dynamical modeling and the update of the planetary data sets used for its adjustment. New determinations for the Sun oblateness including the Lense-Thirring effect, different scenario for the rotation of the Sun core,  and the mass of the Kuiper belt are given in Sect  \ref{sec:LT}, \ref{sec:srot} and \ref{sec:tno} respectively. Comparisons with our former ephemeris, INPOP19a, are also presented in terms of postfit residuals in Sect. \ref{sec:res}. In Sect. \ref{sec:bcsim} we explain how we simulate Earth-Mercury range measurements for the Bepi-Colombo MORE experiment. In Sect. \ref{sec:methodG} and  \ref{sec:method}, is described the approach associating Monte Carlo sampling and least squares adjustment with WRSS  filtering for obtaining new limits for PPN parameters and secular variations of Sun gravitational mass, $\dot{\mu}/\mu$. Finally we give in Sect. \ref{sec:resinpop} the results obtained with INPOP20a for the PPN parameters $\beta$, $\gamma$ and $\dot{\mu}/\mu$ where Sect \ref{sec:resbc} provides the one obtained in including the Bepi-Colombo simulations.

\begin{table}
       \caption{INPOP20a data samples. Column 1 gives the observed planet and information on the type of observations, and Column 2 indicates the number of observations. Columns 3 and 4 give the time interval and the {\it{\textup{a priori}}} uncertainties provided by space agencies or the navigation teams, respectively. Finally, the WRMS for INPOP19a and INPOP20a are given in the last two columns.}
   \begin{tabular}{l c c c l l}
    \hline
    Planet / Type & $\#$ & Period & {\it{\textup{A priori}}} & \multicolumn{2}{c}{WRMS}\\
    & & & uncertainty & INPOP19a  & INPOP20a \\
    \hline
   {\bf{Mercury}} & & & & &\\
Direct range [m]& 462 &  1971.29 : 1997.60 &   900 & 0.95 & 0.95\\
Messenger range [m]& 1096 &  2011.23 : 2014.26 &     5   & 0.82 & 0.82 \\
Mariner range [m]& 2 &  1974.24 : 1976.21 &    100   & 0.37 & 0.42 \\
{\bf{Venus}} & & & & &\\
VLBI [mas]& 68 &  1990.70 : 2013.14 &   2.0  & 1.13 & 1.15 \\
Direct range [m]& 489 &  1965.96 : 1990.07 &   1400    & 0.98 & 0.98 \\
Vex range [m]& 24783 &  2006.32 : 2011.45 & 7.0   & 0.93 & 0.94 \\
{\bf{Mars}} & & & & &\\
VLBI [mas]& 194 &  1989.13 : 2013.86 &   0.3   & 1.26 & 1.26 \\
Mex range [m]& 30669 &  2005.17 : 2017.37 &  2.0  & 0.98 & 1.0175 \\
& & 2005.17 : 2016.37 &  2.0  & 0.97 & 1.02 \\
MGS range [m]& 2459 &  1999.31 : 2006.70 &   2.0   & 0.93 & 0.945 \\
MRO/MO range [m]& 20985 &  2002.14 : 2014.00 &  1.2  & 1.07 & 1.016 \\
Viking range [m]& 1258 &  1976.55 : 1982.87 &    50.0  & 1.0 & 1.0\\
{\bf{Jupiter }}& & & & &\\
VLBI [mas]& 24 &  1996.54 : 1997.94 & 11   & 1.01 & 0.998 \\
Optical RA/Dec [arcsec]& 6416 &  1924.34 : 2008.49 &   0.3  & 1.0/1.0 & 1.02/1.01 \\
Flyby RA/Dec [mas]& 5 &  1974.92 : 2001.00 &  4.0/12.0 & 0.94/1.0 & 0.95/1.01 \\
Flyby range [m]& 5 &  1974.92 : 2001.00 &    2000 & 0.98 & 1.24 \\
Juno range [m]& 14 &  2016.65 : 2019.84 &  14   & 1.35 & 1.02 \\
& 9 &  2016.65 : 2018.68 & 14 & 1.35 & 1.01 \\
{\bf{Saturn}} & & & & \\
Optical RA/Dec [arcsec]& 7826 &  1924.22 : 2008.34 &    0.3 & 0.96/0.87 & 0.96/0.88\\
Cassini & & & & &\\
VLBA RA/Dec [mas]& 10 &  2004.69 : 2009.31 &   0.6/0.3 & 0.97/0.99 & 0.945/0.973 \\
JPL range [m]& 165 &  2004.41 : 2014.38 &  25.0   & 0.99 & 1.033 \\
Grand Finale range [m]& 9 &  2017.35 : 2017.55&  1.0 & 1.71 & 0.8 \\
Navigation [m] & 572 &  2006.01 : 2009.83 & 6.0 & 0.71 & 0.85 \\
TGF range [m] & 42 &  2006.01 : 2016.61 & 15.0 & 1.13 & 1.30 \\
{\bf{Uranus}} & & & & &\\
Optical RA/Dec [arcsec]& 12893 &  1924.62 : 2011.74 &   0.2/0.3  & 1.09 / 0.82 & 1.09 / 0.82 \\
Flyby RA/Dec [mas]& 1 &  1986.07 : 1986.07 &   50/50 & 0.12 / 0.42 & 0.133 / 0.40 \\
Flyby range [m]& 1 &  1986.07 : 1986.07 & 50 & 0.92 & 0.92 \\
{\bf{Neptune}} & & & & &\\
Optical RA/Dec [arcsec]& 5254 &  1924.04 : 2007.88 &   0.25/0.3  & 1.008 / 0.97 & 1.008 / 0.97  \\
Flyby RA/Dec [mas]& 1 &  1989.65 : 1989.65 &  15.0 & 0.11 / 0.15 & 0.11 / 0.15 \\
Flyby range [m]& 1 &  1989.65 : 1989.65 &    2 & 1.14 & 3.6805\\
\hline
    \end{tabular}
    \label{tab:res}
\end{table}

\section{INPOP20a planetary ephemerides}
\label{sec1}
The INPOP20a planetary ephemerides was built with the same data sample as INPOP19a \citep{2019NSTIM.109.....V} but with the addition of 5 Jupiter positions deduced from the Juno perijove PJ19 to PJ23, leading to a coverage of more than 4 years with an accuracy of about 14 meters. Two important modifications have also been brought to the dynamical modeling and are presented in the following.

\subsection{Lense-Thirring effect}
\label{sec:LT}
When comparing the estimations of the Sun oblateness \footnote{this definition corresponds to the gravity field second degree term, -C20}, $J^{\odot}_{2}$, obtained with planetary ephemerides to values obtained by helioseismology \citep{2008A&A...477..657A, 1998MNRAS.297L..76P}, it is important to keep in mind that an additional contribution must be included in order to compare consistent estimates: the effect of the Sun rotation on the space-time metric \citep{1918PhyZ...19..156L}. This effect known as the Lense-Thirring effect has been evaluated to contribute to about 10$\%$ \citep{Hees2015} of the dynamical acceleration induced by the shape of the Sun in General relativity (GRT). With the accuracy of the Bepi-Colombo mission, it is important to include this effect in the INPOP equations of motion. The acceleration induced by the Lense-Thirring effect generated by a central body (at the first post-Newtonian approximation) is given by
\begin{equation}
\vec{a}_{LT} = \frac{(\gamma+1)G}{c^{2}r^{3}} S [ 3 \frac{\vec{k} . \vec{r}}{r^{2}} (\vec{r} \wedge \vec{v}) -  (\vec{k} \wedge \vec{v}) ]
\label{eq:LT}
\end{equation}
where $G$ is the gravitational constant, $c$ the speed of light, $\vec{S}$ is the Sun angular momentum such as $\vec{S} = S \vec{k}$ where $\vec{k}$ is the direction of the Sun rotation pole defined according to the IAU right ascension and declination \citep{2018CeMDA.130...22A}, $\vec{r}$ and $\vec{v}$ are the position and velocity vectors of the planet relative to the central body (here the Sun) and $\gamma$ is the PPN parameter for the light deflection.
Depending the model adopted for the rotation of the Sun core (see Sect. \ref{sec:srot}), one can estimate different values for the amplitude of the Sun angular momentum S, implemented in INPOP and presented in Table \ref{tab:Sun}.
For each value of the Sun angular momentum, an INPOP adjustment is done and  $J^{\odot}_{2}$ is estimated. The $J^{\odot}_{2}$ obtained with INPOP20a in considering the Sun angular momentum from helioseismological measurements \citep{1998MNRAS.297L..76P} is given in the first line of Table \ref{tab:Sun}. This value,  (2.21 $\pm$ 0.01)$\times 10^{-7}$, is very close from the values deduced from SOHO (2.22 $\pm$ 0.009)$\times 10^{-7}$ and GONG (2.18 $\pm$ 0.005)$\times 10^{-7}$ \citep{2008A&A...477..657A}. 
It is also in good agreement with the previous analysis of the same data made by \citep{1998MNRAS.297L..76P} giving as an average estimate between GONG and SOHO, (2.18 $\pm$ 0.06) $\times 10^{-7}$. 
In \citep{2017AJ....153..121P}  estimations for both S and  $J^{\odot}_{2}$ (presented in Table \ref{tab:Sun}) were obtained in considering Messenger tracking data. 
It is important to stress that there is an important correlation (80 $\%$) between S and $J^{\odot}_{2}$  when both estimated in a global planetary fit.
Because of this high correlation, the Sun angular momentum S is not fitted in the INPOP adjustment instead we use the value from \citep{1998MNRAS.297L..76P}. The same choice has been made by \citep{2018NatCo...9..289G} who focus on using Messenger data for constraining Mercury and Earth orbits. Their obtained value of $J^{\odot}_{2}$ is also given in Table \ref{tab:Sun} and is consistent with our estimate as well as with the one of \citep{2017AJ....153..121P} and \citep{1998MNRAS.297L..76P} but not with \citep{2008A&A...477..657A}. 
Finally, with the planetary ephemerides determinations, the PPN parameter $\beta$ and the Sun oblateness $J^{\odot}_{2}$ are usually strongly correlated. A simultaneous estimation of these two quantities is usually very complex or leads to underestimated uncertainties \citep{2015CeMDA.123..325F, 2018NatCo...9..289G}. 
For this reason and as it is now possible to directly related INPOP $J^{\odot}_{2}$ with helioseismological values, we chose to constraint the fitted values of  $J^{\odot}_{2}$  to remain in between the interval of (2.18 $\pm$ 0.06) $\times 10^{-7}$ corresponding to \citep{1998MNRAS.297L..76P} results. We see in Table \ref{tab:Sun} that in the case of GRT (with $\beta$ and $\gamma$ equal to one), this interval is easily respected. The case where GRT is violated is discussed in Sec. \ref{section2}.

\begin{table}
\caption{Sun Angular Momentum and oblateness. Is given in Column 3, the values of Sun $J^{\odot}_{2}$ obtained after fit using the values of the amplitude of the angular momentum given in Column 2. Different models of rotation (identified in Column 1) are used for estimating S. In the first line, are given the results obtained for INPOP20a. See Sect \ref{sec:srot} for the significance of the different rotation hypothesis.}
\begin{tabular}{l l l}
\hline
Type of rotation & S $\times 10^{48}$ & $J^{\odot}_{2}$ $\times 10^{7}$ \\
& g.cm$^{-2}$.s$^{-1}$& \\
\hline
INPOP20a with Pitjers 1998  & 1.90 & 2.218 $\pm$ 0.03 \\
\citep{2017AJ....153..121P}& 1.96 $\pm$ 0.7 & 2.280 $\pm$ 0.06 \\
\citep{2018NatCo...9..289G}& 1.90  & 2.2710 $\pm$ 0.003 \\
\\
Slow rotation &1.896 & 2.208 $\pm$ 0.03 \\
uniform rotation at 435 nHz & 1.926 & 2.210 $\pm$ 0.03 \\
Fast rotation &1.976 & 2.213 $\pm$ 0.03 \\
Very fast rotation & 1.998 & 2.214 $\pm$ 0.03  \\
\hline
Pijpers (1998) & 1.90 $\pm$ 1.5 & 2.180 $\pm$ 0.06  \\
Antia et al. (2008) & 1.90 $\pm$ 1.5 & 2.2057 $\pm$ 0.007 \\
\hline
\end{tabular}
\label{tab:Sun}
\end{table}

\begin{figure}
    \centering
    \includegraphics[scale=0.38]{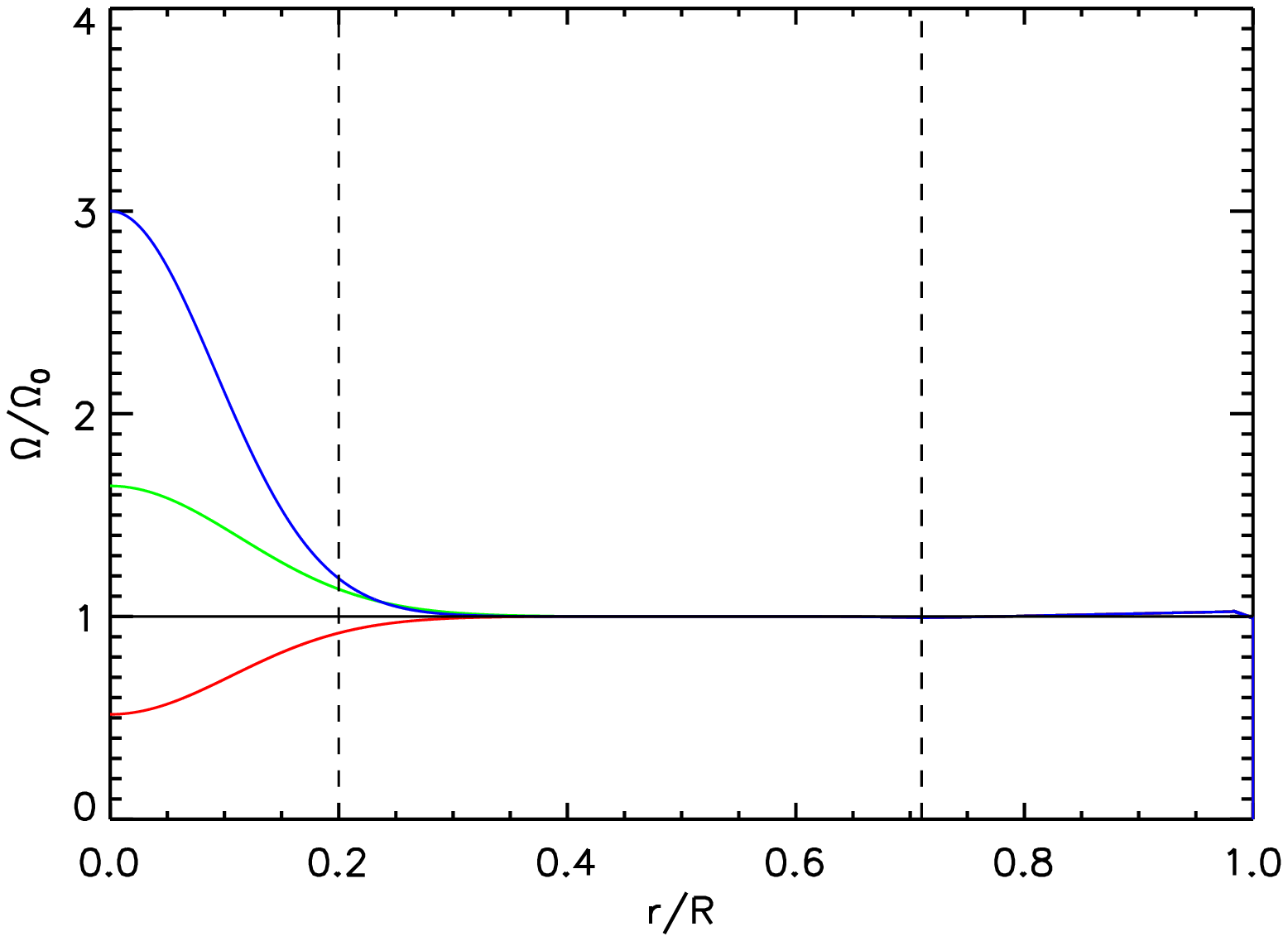}\includegraphics[scale=0.38]{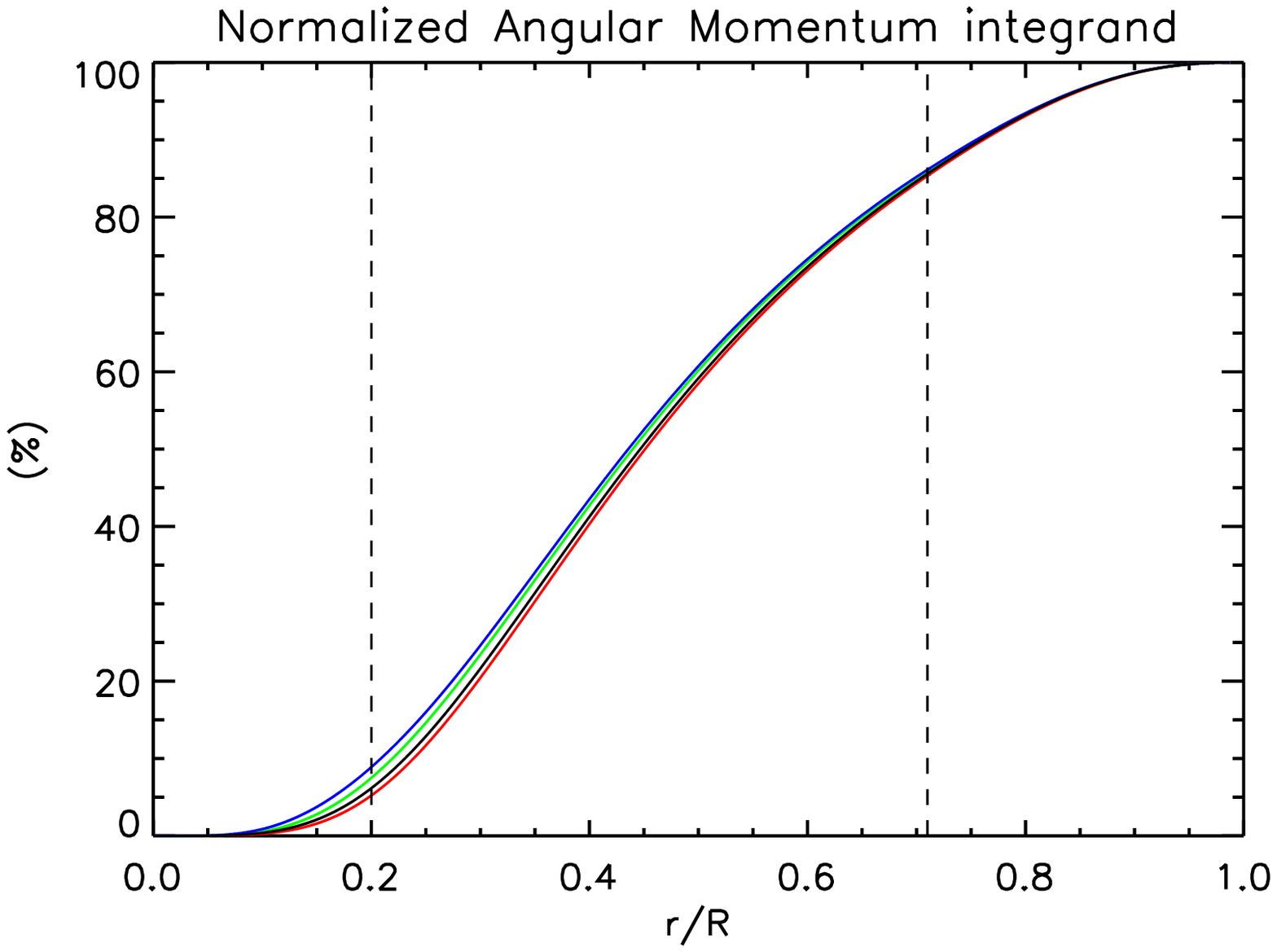}
    \caption{(Left Panel) The four rotation profiles as functions of depth used to computed the angular momentum of the Sun normalized by $\Omega_0 = 435\,{\rm nHz}$ : very fast rotation (blue), fast rotation (green), slow rotation (red), and uniform rotation (black). The dashed lines represent the approximate boundaries of the solar core at $r \sim 0.2\,R_\odot$ and the position of the tachocline, i.e. the transition between the radiative and the convection zones. (Right panel) The corresponding angular momentum integrands in Eq. \ref{eq:momemtum} as functions of depth.}
    \label{fig:Sun} 
\end{figure}

\subsection{Sun core rotation}
\label{sec:srot}
We use the INPOP planetary ephemerides to constrain solar core rotation. 
In our global solution of the planetary ephemeris, solar rotation is present through the gravitational $J^{\odot}_{2}$ and the Lense-Thirring effects. While the first effect is coming as a solution of the fit of these ephemerides, the latter needs the knowledge of the solar angular momentum calculated using a calibrated solar model:  
\begin{equation}
S = \frac{1}{2} \int_0^{R_\odot} r^2 dm \int_{-1}^1 (1-\cos^2\theta)\,\Omega(r,\theta)\,d\cos\theta,
\label{eq:momemtum}
\end{equation}
where $dm=4\pi \rho r^2 dr$ is the mass fraction. The rotation profile is splitted in two parts :  $\Omega (r,\theta) = \Omega_{\rm core} (r) + \Omega_{\rm helio}(r,\theta)$. We adopt the solution $\Omega_{\rm helio}(r,\theta)$ proposed in \citet{2001A&A...377..688R} inferred from helioseismology. 

For $\Omega_{\rm core}$, we assume only a radial dependence. The profile of the rotation is assumed to have the parametric form $\Omega_{\rm core}/2\pi = {\rm K\, exp(-r^2/r_c^2)} $, where ${\rm K}$ is a constant to adjust and $r_c \sim 0.15\,R_\odot$ characterizes the extension of the core. These coefficients are adjusted to reproduce the rotation rates inferred from helioseismology and shown in Fig. 16 of \citet{2003LNP...599...31D}. The result of the fit is shown in Fig. \ref{fig:Sun} together with the case of a very fast rotation rate, as proposed in Fossat et al. For the present analysis, we consider 4 cases : very fast (GOLF, \cite{2017A&A...604A..40F}), fast (GOLF, \cite{1998ESASP.418..329R}), slow (Bison,  \cite{1999MNRAS.308..405C}), and uniform. 
The values of angular momenta and corresponding $J^{\odot}_{2}$ are shown in Table \ref{tab:Sun}. The different core rotations  change the total angular momentum by about 5\% at most between the two extreme cases (very fast and slow core rotations). The reason of such small impact of the core rotation is due to the $r^2$ dependence in Eq. \ref{eq:momemtum} as seen in Fig. \ref{fig:Sun}. Most of the contribution of the angular momentum ($80\%$) is coming from the radiative zone and the core has a small contribution. We have also tested the impact of differential rotation of the convection zone, as inferred by helioseismology, i.e. the radial and latitudinal dependencies, compared to the  case of uniform solid rotation. The difference is very small, i.e. $0.02\%$. 
Using these different angular momenta to account for the Lense-Thirring effect in our global ephemerides fit, we extract the corresponding gravitational $J^{\odot}_{2}$, as shown in Tab. \ref{tab:Sun}. Our values found by fitting planetary ephemerides are in good agreement with those inferred from helioseismology \citep{1998MNRAS.297L..76P, 2008A&A...477..657A} with a value close to $2.2 \times 10^{-7}$. We emphasize that our value using the very fast core (e.g. following Fossat et al.) is smaller than the $J^{\odot}_{2} \approx 2.6 \times 10^{-7}$ found in \citet{2019ApJ...877...42S}. The reason of this difference is due to their large extent of the fast rotating core.  Our differences in $J^{\odot}_{2}$ coming from the different core rotations are much smaller than our error bars, which prevents us to disentangle these core rotations with the current planetary ephemerides. We will see that with the inclusion of the Bepi-Colombo simulations (BC), this conclusion could be different (Sect. \ref{sec:J2bc}).

\subsection{Trans-Neptunian objects}
\label{sec:tno}

\begin{figure}
    \centering
    \includegraphics[scale=0.3]{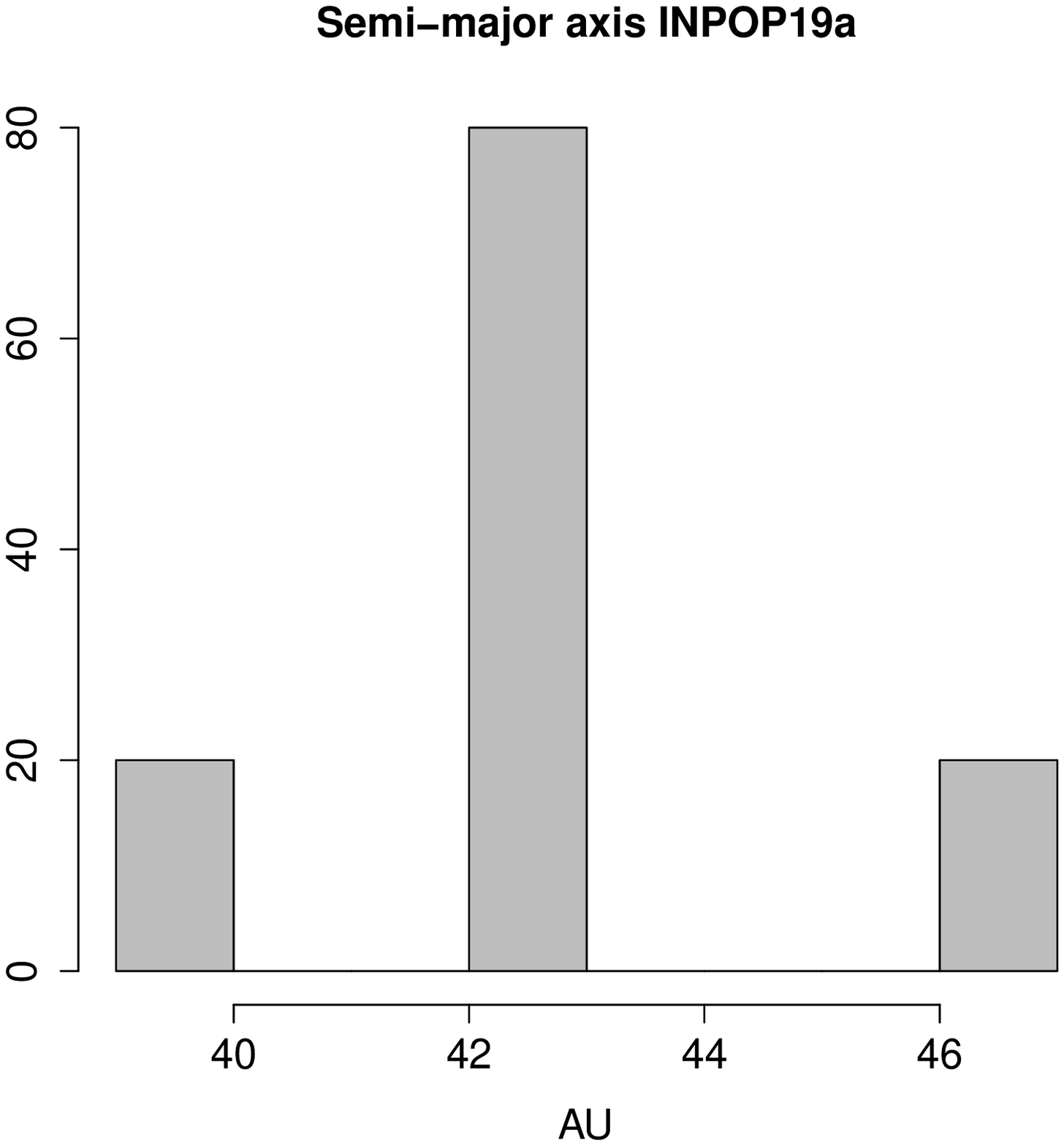}\includegraphics[scale=0.3]{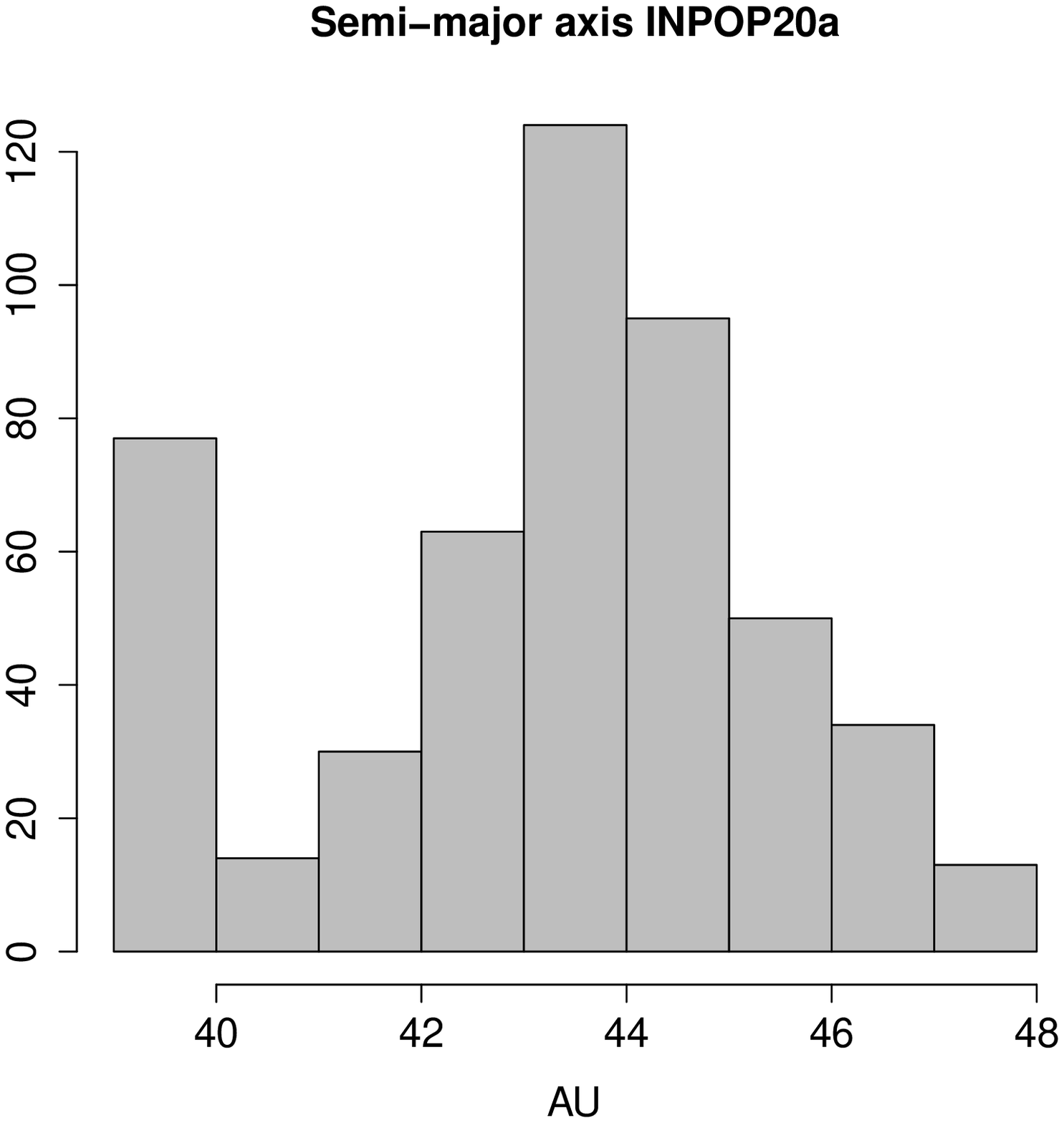}\includegraphics[scale=0.3]{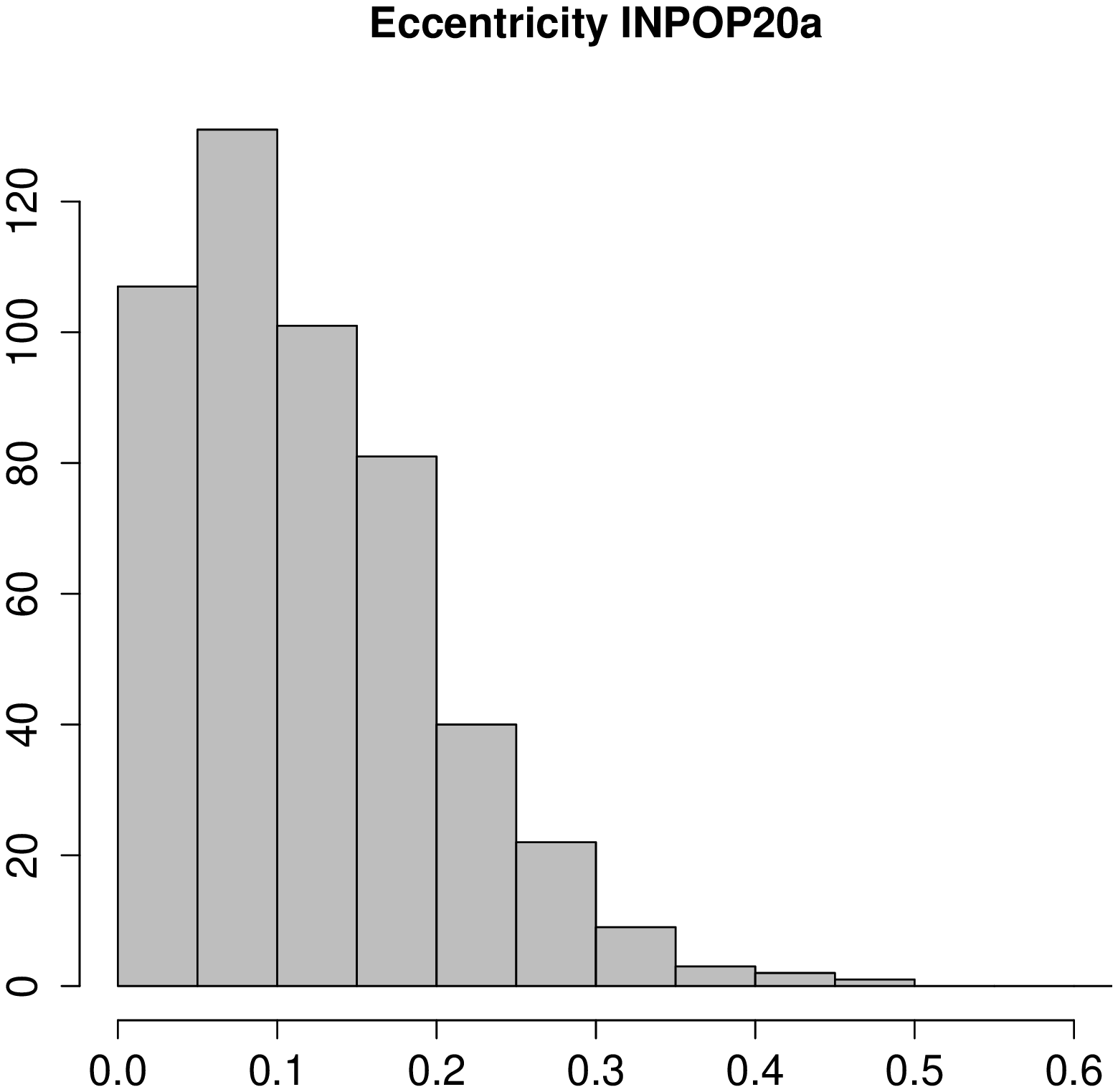}
    \caption{Semi-major axis and eccentricity distributions: Comparison between the INPOP19a TNO modeling and one selection of 500 orbits randomly chosen in the Astorb database. }
    \label{fig:modeletno2}
\end{figure}

In INPOP19a, a modeling based on three circular rings representing the perturbations of Trans-Neptunian objects (TNO) located at 39.4, 44.0 and 47.5 AU has been introduced and outer planet orbits have been clearly improved, especially Saturn orbit \citep{2019NSTIM.109.....V, 2020A&A...640A...7D}. However, with this circular ring modeling, the impact of the eccentricities of the TNO orbits was not included in the computation of the perturbing accelerations. The global mass of these rings appears also to be too important in comparison with theoretical estimations (see \citep{2020A&A...640A...7D} for the full discussion). As TNO orbits tend to be more eccentric compared to main belt asteroid orbits, we implement an alternative representation by considering directly observed orbits extracted from the Astorb database \citep{2018DPS....5040808M}. 
In order to limit the number of objects to consider for not increasing too much the time of computation, on the total of 2225 objects with semi-major axis between 39.3 and 47.6 AU, we operated random selections of 500 of them that we integrated as individual objects with the same mass spread over the 500. Thanks to this approach the representation of the TNOs is more realistic, in particular, regarding the distributions in eccentricities and in semi-major axis (see Fig. \ref{fig:modeletno2})  without increasing too much the integration time. For each random sampling of 500 objects a full fit was operated in adjusting the global mass of the 500, in addition to the regular planetary ephemeris parameters. Results being very similar from one random selection to another, one selection was chosen arbitrary for the rest of this study.
After fit, the global mass for 500 TNOs is found to be (1.91 $\pm$0.05)$\times 10^{23}$ kg which corresponds to (0.031 $\pm$ 0.001) M$_{\Earth}$. This mass is about two times smaller than the one proposed by \citep{2020A&A...640A...7D}. This difference can be explained by the differences in the dynamical modeling between this work and \citep{2020A&A...640A...7D}. While \citep{2020A&A...640A...7D} used circular rings, we include here a real distribution of orbits with various eccentricities as one can see on Fig. \ref{fig:modeletno2}. 

\subsection{Postfit residuals and comparisons with INPOP19a}
\label{sec:res}

One can find on Table \ref{tab:res} the  weighted root mean squares (WRMS) for the INPOP19a and INPOP20a postfit residuals together with a brief description of the corresponding data sets (type of observations, number of observations, time coverage). The WRMS is defined as $WRMS=\sqrt{\sum_{i=1}^{N} \frac {(O_i-C_i)^2}{\sigma_i ^ 2}}$, where $(O_i-C_i)$ is the postfit residual for the observation $i$, $\sigma_{i}$ is the {\it{a priori}} instrumental uncertainty of the observation $i$ given in Column 4 of Table \ref{tab:res}.
Where for the inner planets, the differences between INPOP19a and INPOP20a are not clearly visible, the improvement is more effective for outer planets, especially for Jupiter and Saturn.
For Jupiter, the addition of 5 new perijove obtained up to P23 improves the residuals from about 18~m with INPOP19a to 14~m with INPOP20a. 

\begin{figure}
\centering
    \includegraphics[scale=0.4]{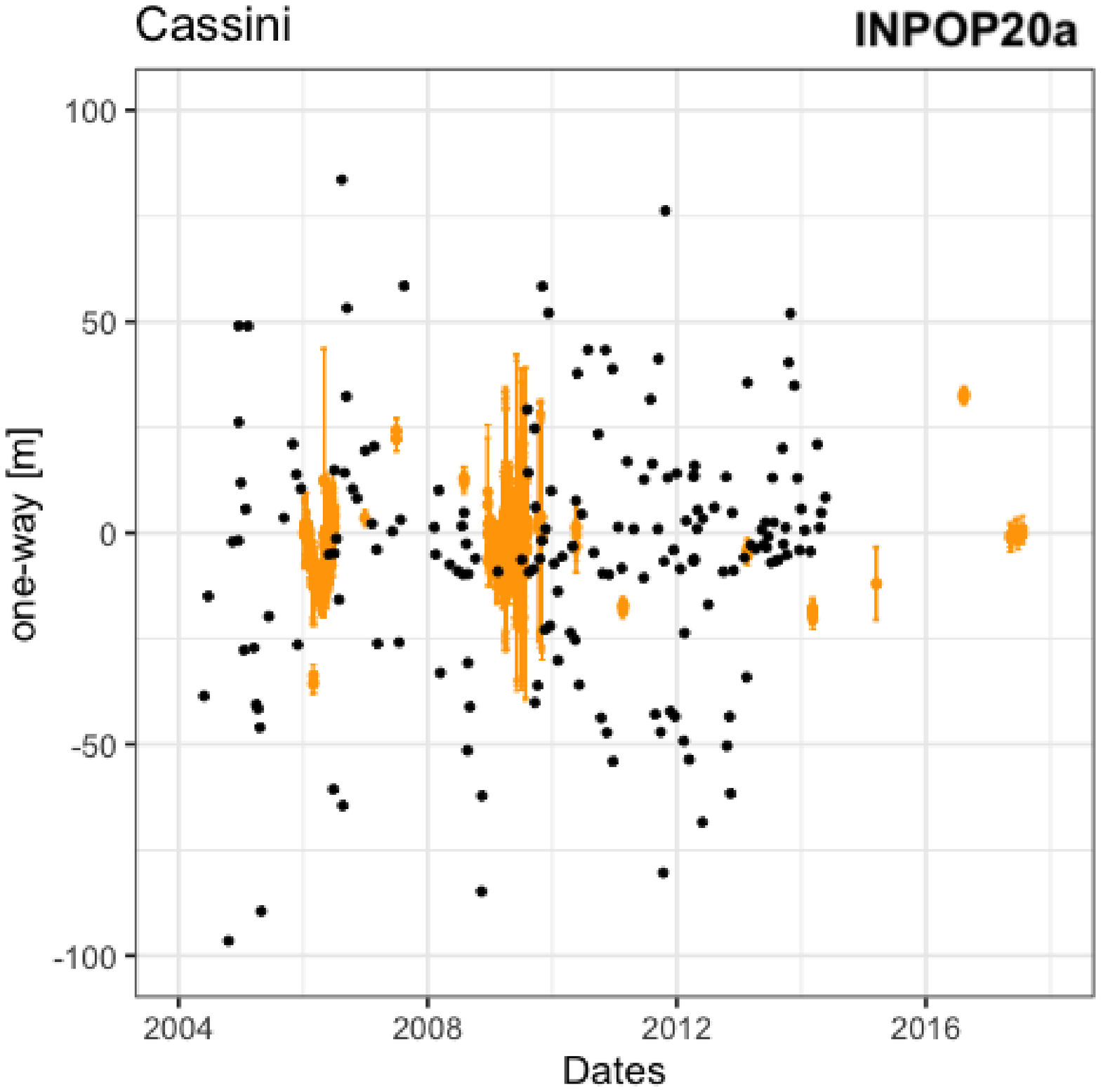}\includegraphics[scale=0.4]{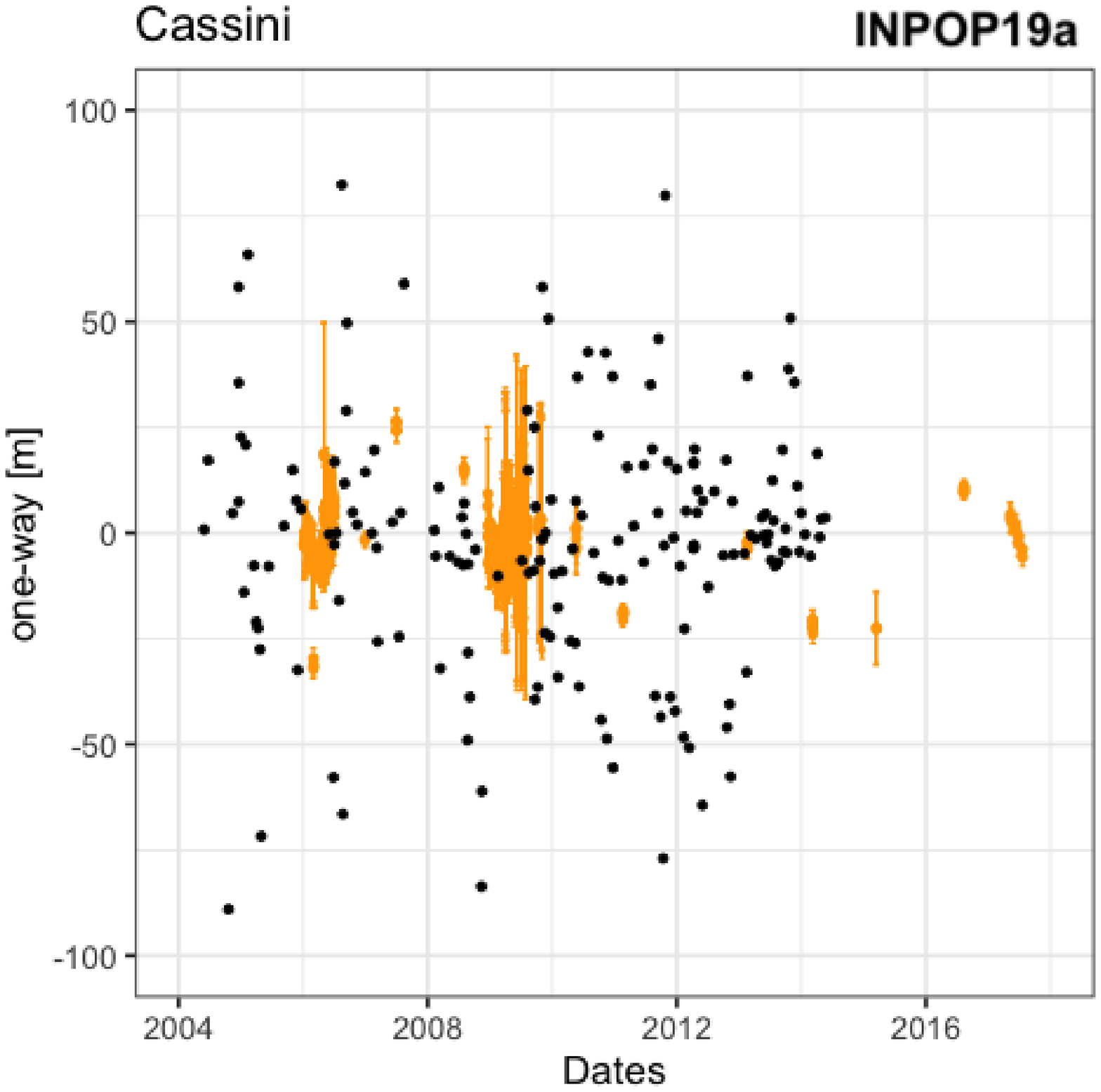}
    \caption{Cassini postfit residuals for INPOP19a and INPOP20a. The black dots are residuals obtained for JPL range sample when the orange dots indicate the residuals obtained for the Grand Finale (in 2017), the Navigation and the TGF range samples. For a more complete description of these sample, see \citep{2020A&A...640A...7D}.}
    \label{fig:cass}
\end{figure}

For Saturn, one can note the significant improvement for Cassini samples, in particular for the Grand Final residuals obtained in 2017 for which the INPOP20a residuals is about 2.4 times smaller than the INPOP19a one. This result is a direct consequence of the introduction of the new TNO ring modele presented in Sect. \ref{sec:tno} and is linked to the removal of a secular trend clearly visible in INPOP19a residuals but not in INPOP20a (see Fig. \ref{fig:cass}).

\subsection{Bepi-Colombo simulations}
\label{sec:bcsim}
\begin{figure}
    \centering
    \includegraphics[scale=0.5]{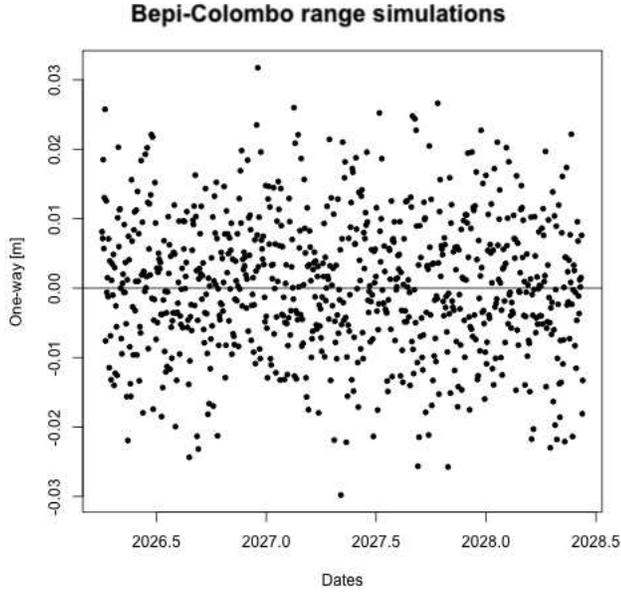}
    \caption{Bepi-Colombo 1-cm simulated residuals obtained in using INPOP20a as reference ephemeris in GRT and an 8-hour temporal resolution. }
    \label{fig:bcsim}
\end{figure}

In using INPOP20a as a reference planetary ephemerides, we have simulated possible range bias for the Bepi-Colombo mission between Mercury and the Earth. 
These simulations are used for estimating the impact for the INPOP construction of using very accurate Mercury-Earth distances as they should be obtained by the radio science MORE experiment. Based on the assumption that the radio tracking in KaKa-Band keeps the 1~cm accuracy that has been monitored during the commissioning phase of the Bepi-Colombo mission in 2019 and 2020 \citep{2021SSRv..217...21I}, we suppose a daily acquisition of range tracking data \citep{thor20} during a period of 2.5, from 2026 to 2028.5. The simulated residuals obtained in using INPOP20a as reference ephemerides are plotted in Fig \ref{fig:bcsim}. 
These GRT residuals provide a reference against which can be tested the epheremides integrated with non-GRT parameters (i.e., PPN $\beta \ne  \gamma \ne 1$ and $\dot{\mu}/\mu \ne 0$).
The capability of these alternative ephemerides to have a good fit with the GRT simulated observations will tell us what constraints can be obtained on the PPN parameters.
In the same manner we have tested the sensitivity of the BC simulations to any change in the values of the Sun angular momentum (see Sect \ref{sec:J2bc}). These simulations are then added to the INPOP20a data sample for building a new reference GRT ephemeris (see Sect. \ref{sec:cdi}) and new alternatives ephemerides in non-GRT (see Sect. and \ref{sec:bg_bc} \ref{sec:mu_bc}).

\section{ GRT violations with INPOP20a and  Bepi-Colombo simulations}
\label{section2}

\subsection{Method}
\label{sec:methodG}

By construction, the planetary ephemerides cannot disentangle the contribution of the PPN parameters $\beta$, $\gamma$ and the Sun oblateness $J^{\odot}_{2}$ \citep{2015CeMDA.123..325F, Bernus2020b}. The introduction of the Lense-Thirring effect helps for individualize the signature induced by PPN $\gamma$ but correlations between these parameters stay high. This is the reason why a direct adjustment of these three parameters together in a global fit leads to highly correlated determinations and under-estimated uncertainties. One way to overcome this issue is to fix one of these contributors, for example in fixing the $\gamma$ value to the one estimated by \citep{bertotti2003nat} with the Cassini experiment in 2003. However, as with the Bepi-Colombo mission, far more accurate constraints are planned to be obtained with the same solar conjunction techniques \citep{2017CQGra..34g5002I}, we decide not to fix $\gamma$ but to add helioseismological limits for the Sun $J^{\odot}_{2}$ (see Sect \ref{sec:LT}). These  thresholds are  applicable when the Lense-Thirring effect is included in the dynamical modeling. Additionally, as explained in Sect \ref{sec:LT}, in the GRT case when no limits are applied to the $J^{\odot}_{2}$ determinations, INPOP20a gives a very consistent value, included in the uncertainties of the heliosismology. 
We note that the helioseismology limits are obtained based on the analysis of time variations of the Sun angular momentum and its kinetical energy in the Newtonian framework. Even if a strict approach would require a complete reanalysis of the helioseismology measurements in a non GRT-frame, one can expect a negligeable effect \citep{2016JGeod..90.1345S}.
As in \citep{2015CeMDA.123..325F}, we introduce the parameter $\dot{G}/G$ through the secular variations of the gravitational mass of the Sun $\dot{\mu}/\mu$ with the equation
\begin{equation}
\dot{\mu}/\mu = \dot{G}/G + \dot{M_{\odot}}/M_{\odot}
\end{equation}
where $\dot{M_{\odot}}/M_{\odot}$ is the Sun mass loss. This quantity is fixed by \citep{2015CeMDA.123..325F} such as
\begin{equation}
\frac{\dot{M_{\odot}}}{M_{\odot}} = { {(-0.92 \pm 0.61) \times 10^{-13} \quad (3\sigma)}} \, \textrm{yr}^{-1} 
 \label{mudotdef2}
\end{equation}
$\dot{G}/G$ is also accounting for the update of the masses of the planets and asteroids at each step of the integration of the body equations of motion as well as in the time-scale transformation TT-TDB and the correction of the Shapiro delay in the range computation. At each step $t$ of the numerical integration  of the INPOP equations of motion, we then estimate :
\begin{eqnarray}
M_{\odot}(t) &=& M_{\odot}(J2000) + (t-J2000) \times \dot{M_{\odot}} \\
G(t) &=& G(J2000) + (t-J2000) \times \dot{G}\\
\mu(t)&=& G(t)  \times M_{\odot}(t) 
\label{eq:mudot}
\end{eqnarray}
where $M_{\odot}(J2000), G(J2000)$ are the mass of the sun and the constant of gravitation at the date J2000 and $(t-J2000)$ is the time difference between the date of the integration $t$ and J2000 in years.
The value of the Sun gravitational mass $\mu(t)$ corresponding to the date of the observation $t$ is computed with Equation \ref{eq:mudot} and is then re-introduced in the Shapiro delay equation (8-38) given in \citep{Moyer2000}.
In this context, the strategy chosen for this study is the same as in \citep{2015CeMDA.123..325F, bernus2019, 2020PhRvD.102b1501B}: we built full planetary ephemerides by integrating and adjusting to observations Einstein-Imfeld-Hoffman equations of motion for planetary orbits, timescale transformation and Shapiro delay computation  \citep{Moyer2000, 2009A&A...507.1675F} together with the Lense-Thirring effect (Eq. \ref{eq:LT}) and time varying  G (Eq. \ref{eq:mudot}). For the non-GRT parameters (PPN $\beta \ne 1$, $\gamma  \ne 1$ and $\dot{\mu}/{\mu} \ne 0$), we take samples of random values following uniform distributions. Table \ref{tab:unif} gives the intervals used for these uniform distributions. We have chosen random distributions instead of regular grids because we plan to use these results in a Bayesian context for a forthcoming study.
Finally, these intervals were chosen in order to encompass the larger published limits \citep{2015CeMDA.123..325F}. For the BC simulation, the intervals were optimized according to the sensitivity of the simulated observations but still encompass the limits proposed by the literature in the frame of the BC mission \citep{2018Icar..301....9I,2020CQGra..37i5007D}.
In this work, for filtering out the ephemerides built with the non GRT parameters (alternative ephemerides), we consider a study of  the WRSS  distribution.

\begin{table}[]
    \centering
    \begin{tabular}{c|c c }
    \hline
   Parameters & INPOP20a & with BC \\
     \hline
        $\beta$ & $\pm 20 \times 10^{-5}$ & $ \pm 20 \times  10^{-5}$ \\
        $\gamma$ & $ \pm 20 \times 10^{-5}$ & $ \pm 1 \times 10^{-5}$ \\
        $\dot{\mu}/\mu$ & $ \pm 6 \times 10^{-13}  \, \textrm{yr}^{-1}$ & $ \pm 1 \times 10^{-13} \, \textrm{yr}^{-1}$ \\
        \hline
    \end{tabular}
    \caption{Intervals for the uniform distributions for PPN parameters $\beta$ and $\gamma$ and $\dot{\mu}/\mu$.}
    \label{tab:unif}
\end{table}

\subsection{Construction of the Weighted Residual Sum of Squares distribution}
\label{sec:method}
\begin{figure}
    \centering
    \includegraphics[scale=0.5]{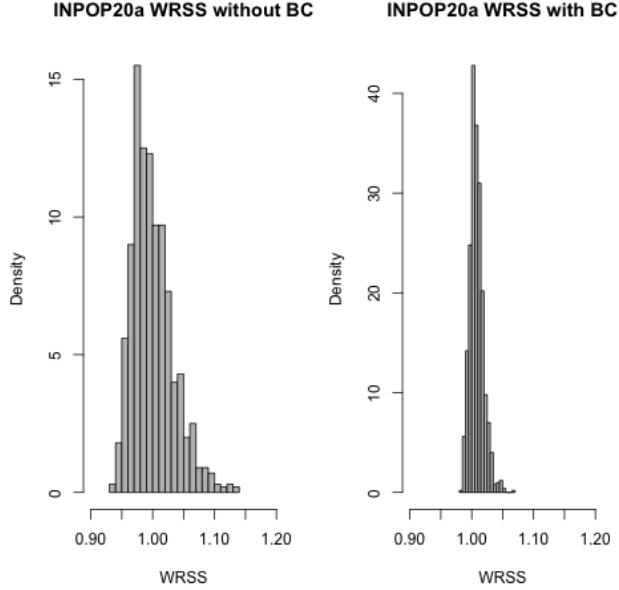}
        \caption{Distribution of INPOP20a WRSS  including instrumental noise without BC simulations (left-hand side) and with BC simulations (right-hand side).}
    \label{fig:chi2woBC}
\end{figure}

Before considering GRT violations, we study the instrumental noise variability of the INPOP adjustment.
To do so, we use Monte Carlo simulations where we generate fake observations obtained by adding a Gaussian noise to the true observations. The standard deviation of the noise added to each observation is taken as the value of the INPOP residual  for the corresponding observation.
We operate 1000 samplings and for each of them we refit the INPOP ephemerides and we compute the Weighted Residual Sum of Squares (WRSS ) as followed:

 \begin{equation}
WRSS  = \frac{1}{N} \sum_{i=1}^{N} \frac{((O-C)_{i})^{2}}{\sigma_{i}^{2}}
\label{eq:chi2}
 \end{equation}

where $(O-C)_{i}$ is the difference between the observation O and the observable C computed with INPOP (postfit residual) for the observation $i$, $\sigma_{i}$ is the {\it{a priori}} instrumental uncertainty of the observation $i$ and $N$ is the number of observations. 
We obtain a experimental WRSS  distribution as presented in Fig. \ref{fig:chi2woBC}. From this empirical distribution, we can estimate the probability of a postfit WRSS  to be explained by the instrumental uncertainties. 
We derive the quantiles corresponding to the 3-$\sigma$ of the WRSS  distribution after fitting a log-normal profile to this latest and we can estimate a confidence interval $[$WRSS $_{max}$:WRSS $_{min}]$ that contains 99.7$\%$ of the distribution. We used these WRSS $_{max}$ and WRSS $_{min}$ as thresholds for the selection of alternative ephemerides (estimated with non-GRT parameters) compatible at 99.7$\%$ with the observations. With the INPOP20a data sampling, this leads to the definition of an interval of WRSS  of about $\pm 0.09$ around the INPOP20a WRSS .
We proceed in the same manner with the Bepi-Colombo (BC) simulations. We add BC simulations to the INPOP20a data sample in taking 1$~$centimeter as instrumental uncertainty. Fig. \ref{fig:chi2woBC} gives the distribution  of the WRSS  including the BC simulations. The obtained profile for the WRSS  distribution is clearly more narrow compared with the one obtained without. This may indicate the improvement of the fit quality and consequently, the increase of the constraint for the tested parameters. In considering the quantiles corresponding to the 3-$\sigma$ WRSS  distribution, we deduce the WRSS $_{min}$ and WRSS $_{max}$ to be used for selecting alternative ephemerides. The WRSS  interval is then of $\pm 0.03$ around the WRSS  of INPOP20a including BC.

\begin{table}
\caption{PPN $\beta$ and $\gamma$ confidence intervals given at 99.7 $\%$ of the reference WRSS  distribution, with and without Bepi-Colombo simulations. The first column indicates the method being used for obtaining the results given in Columns 4 (with INPOP20a) and 6 (with INPOP20a and BC simulations).LS stand for Least squares, MC for Monte Carlo and GA for Genetic Algorithm.}
\label{tab:sres}
\begin{tabular}{l | c c c c}
\hline
& \multicolumn{1}{c}{$(\beta - 1)$} & \multicolumn{1}{c}{$(\gamma - 1)$} &   \multicolumn{1}{c}{$\dot{\mu}/\mu $}& $J^{\odot}_{2}$\\
& \multicolumn{1}{c}{$\times 10^{5}$} & \multicolumn{1}{c}{$\times 10^{5}$} &  \multicolumn{1}{c}{$\times 10^{13}$ yr$^{-1}$} & $\times 10^{7}$ \\
\hline
3-$\sigma$ WRSS  INPOP20a &  -1.12 $\pm$ 7.16  &   -1.69 $\pm$ 7.49  &    -1.03 $\pm$ 2.28 & 2.206 $\pm$ 0.03\\
non-GRT LS INPOP20a  &   -1.9 $\pm$  6.28 &    2.64 $\pm$  3.44  &   -0.37 $\pm$  0.32 & 2.165 $\pm$ 0.12\\
\\
3-$\sigma$ WRSS  INPOP20a + BC &  0.32 $\pm$ 5.00 &  0.09 $\pm$ 0.40 &   -0.19 $\pm$ 0.19  & 2.206 $\pm$ 0.009 \\
non-GRT INPOP20a + BC  &   $\pm$  1.06 &    $\pm$  0.23 &  $\pm$  0.01  &  $\pm$ 0.013\\
\hline
\hline
\\
\citep{2015CeMDA.123..325F} LS 3-$\sigma$ &  -6.7 $\pm$  6.9 & -0.8 $\pm$  5.7  &  -0.50 $\pm$  0.29 & 2.27 $\pm$ 0.25 \\ 
\citep{2015CeMDA.123..325F} MC &  -0.8 $\pm$ 8.2 &  0.2 $\pm$ 8.2 &      -0.63 $\pm$ 1.66 &  1.81 $\pm$ 0.29\\
\citep{2015CeMDA.123..325F} GA & 0.0 $\pm$ 6.9 &  -1.55 $\pm$ 5.01 &      -0.43 $\pm$ 0.74 & 2.22 $\pm$ 0.13\\
\hline
\end{tabular}
\end{table}

\begin{figure}
    \centering
\includegraphics[scale=0.5]{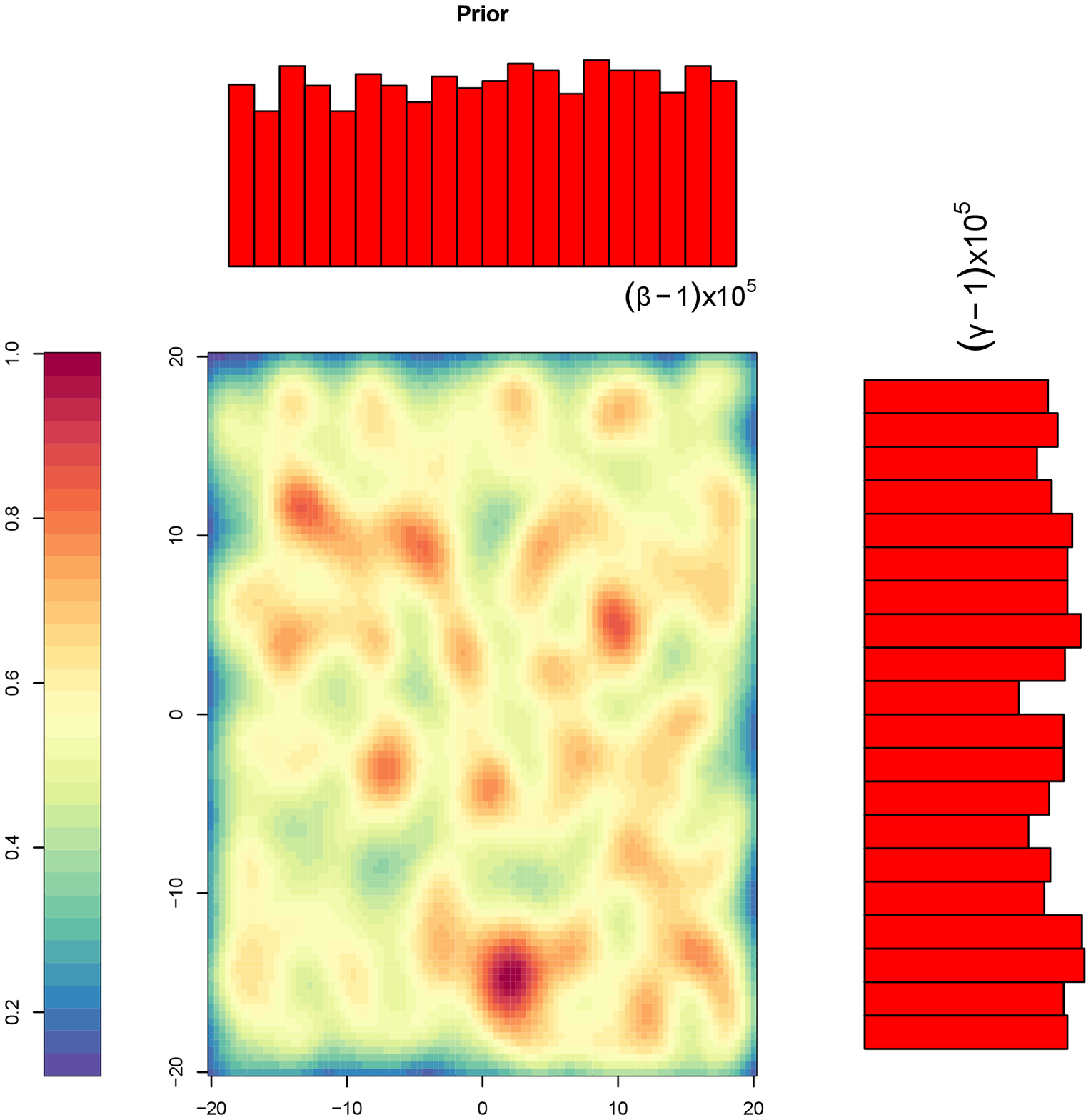}\includegraphics[scale=0.5]{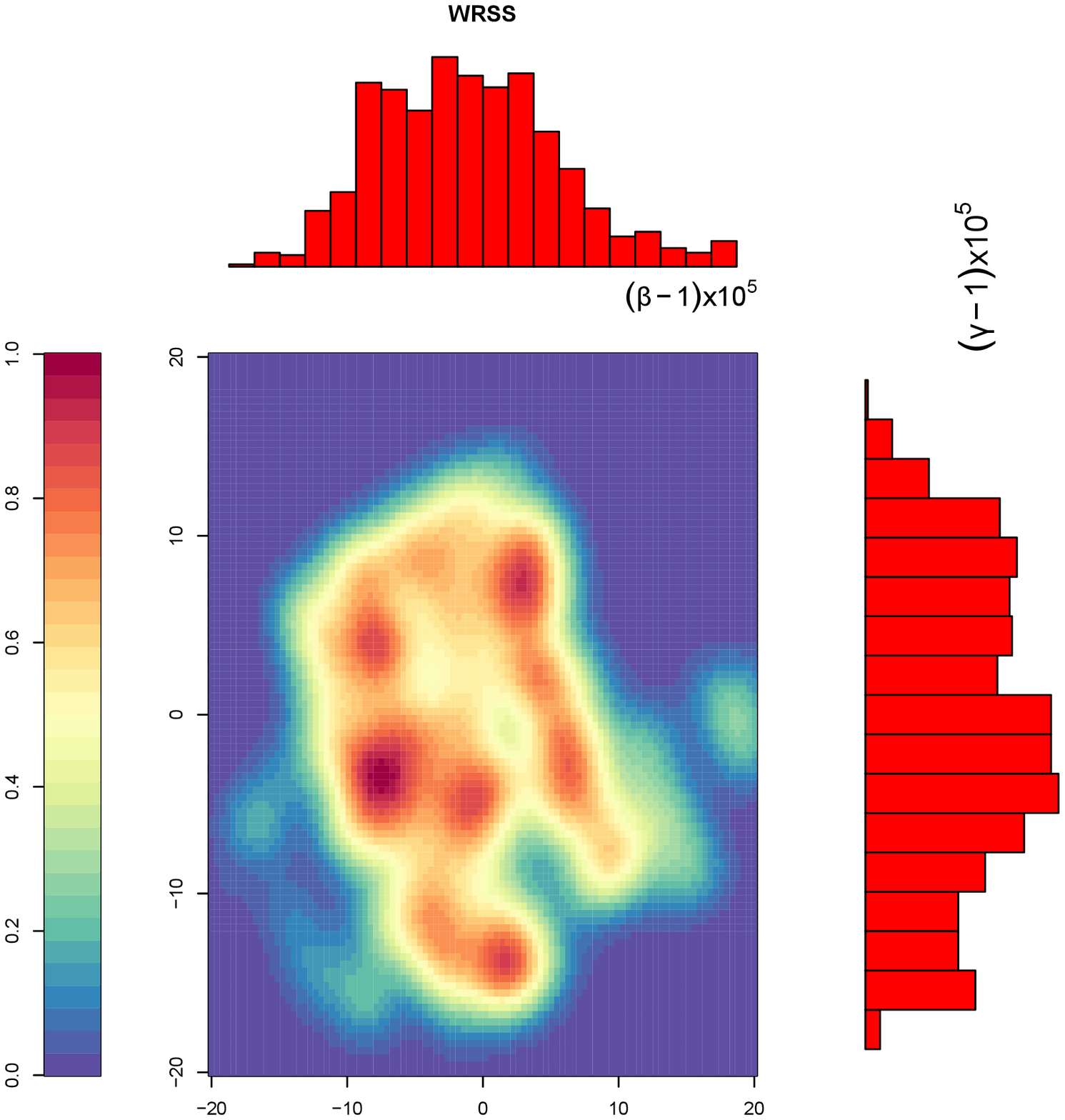}
        \caption{2D-histograms of PPN $\beta$ and $\gamma$. The first left-hand side plot shows the uniform distribution of the prior when the middle histogramms give the distribution of the selected values of $\beta$ and $\gamma$ according to the 3-$\sigma$ INPOP20a WRSS  distribution without considering the Bepi-Colombo simulations. The color-scale indicates the normalised probability.}
    \label{fig:2DwoBC}
\end{figure}

\section{Results}

\subsection{with INPOP20a}
\label{sec:resinpop}
\subsubsection{PPN parameters $\beta$ and $\gamma$}
\label{sec:resppn20a}
3800 runs are estimated with values of PPN $\beta$ and $\gamma$ different from unity, as described in Sect .\ref{sec:methodG}. For each of these runs, alternate planetary ephemerides are integrated and fitted over the same data sample as INPOP20a in an iterative process, equivalent to the one used for INPOP20a construction. When the iterations converged, the WRSS  is compared to the INPOP20a WRSS  distribution as discussed in Sect. \ref{sec:method}. The selection described above is then operated according to the estimated WRSS  and only 23$\%$ of the runs have been kept.
In Fig. \ref{fig:2DwoBC}, are plotted the 2-D histogram for the initial (uniform) distributions of PPN parameters $\beta$ and $\gamma$ together with the 2-D histogram of the selected  ephemerides,  compatible at 3-$\sigma$ with the INPOP WRSS  distribution, without BC simulations. Table  \ref{tab:sres} gives the deduced intervals for the two quantities based on the WRSS  filtering. These intervals are all compatible with GRT and can be compared with the one obtained by direct least square fit (non-GRT LS), based on the INPOP20a data sample. These results are obtained by adding the non-GRT parameters to the full INPOP20a adjustment together with the 402 other parameters (including the mass of the Sun and its oblateness, constrained by helioseismology values). As for the WRSS  filtering, these estimations are also consistent with GRT. Nevertheless, there are major correlations obtained from LS covariance analysis, between $\beta$, $\gamma$ and the other parameters of fit. In particular we note 75$\%$ of correlation between $\beta$ and $J^{\odot}_{2}$ and between  $J^{\odot}_{2}$ and $\gamma$   as well as between $\gamma$ and the semi-major axis of the inner planets. 
As a first consequence of these correlations, in the global fit including $\beta$ and $\gamma$, the value of the $J^{\odot}_{2}$ decreases significantly and could escape from the \citep{1998MNRAS.297L..76P} interval considering the LS uncertainties. 
The WRSS  of this non-GRT fit  is in the 3-$\sigma$ quantiles of the INPOP20a WRSS  distribution, showing that the WRSS  filtering is more conservative than a direct least squares determination. This is also clearly visible when we compare the intervals for $\beta$ and $\gamma$ for the selected alternative ephemerides according to WRSS  filtering and the LS intervals given at 3-$\sigma$: the WRSS  intervals are systematically larger than the LS results.
In Table \ref{tab:sres}, we give also the average of the $J^{\odot}_{2}$ values fitted during the adjustment of the selected $3-\sigma$ WRSS  alternative ephemerides. Without BC simulations, only 12$\%$ of the ephemerides have $J^{\odot}_{2}$ values reaching the heliocentric boundaries. In considering all the selected alternative ephemerides, we obtain a value of $J^{\odot}_{2}$ of $(2.206 \pm 0.03) \times 10^{-7}$. This interval is consistent with the one obtained by LS fit in GRT presented in Table \ref{tab:Sun}. With the direct LS adjustment including non-GRT parameters, we obtain a  3-$\sigma$ uncertainty of about 0.12$\times 10^{-7}$, which is 4 times greater than the 3-$\sigma$ uncertainty obtained for the direct LS fit in GRT or than the dispersion of the fitted $J^{\odot}_{2}$ for the 3-$\sigma$ WRSS selected alternative ephemerides. This comparison shows, as expected, that the  WRSS  filtering allows to obtain less correlated determinations for the $J^{\odot}_{2}$ compared to direct LS adjustment including non-GRT parameters. 
Additionally, one can note that the LS uncertainties obtained with INPOP20a are slightly smaller than the LS estimations obtained by \citep{2015CeMDA.123..325F}, also provided  in Table \ref{tab:sres} (line 4).  In particular, $\gamma$ appears to be more accurately constrained relatively to $\beta$ in the present study than in \citep{2015CeMDA.123..325F}. This can be explained by the correlation between $\beta$ and $\gamma$ which was of 51$\%$ in  \citep{2015CeMDA.123..325F},  falling at about 25$\%$ in this work. This decrease in the correlation is induced by the Lense-Thirring acceleration, introducing an additional constraint on $\gamma$, independently from $\beta$.
In comparisons with \citep{2015CeMDA.123..325F} and the results obtained with a genetic algorithm (labelled "GA" in Table \ref{tab:sres}) and the Monte Carlo runs (labelled "MC" in Table \ref{tab:sres}), one can be surprised that the limits for the $\beta$ and $\gamma$ intervals have not been more reduced as more accurate planetary tracking observations were used for the construction of INPOP20a. 
Several reasons can be proposed.
Firstly, as already mentioned, the reference ephemerides used in \citep{2015CeMDA.123..325F}, INPOP15a, has built using a different dynamical modeling (without Lense-Thirring effect nor TNO ring). The INPOP15a data sample did also not account for Juno data nor Cassini recent re-analysed observations. Secondly the selection criteria used in \citep{2015CeMDA.123..325F} were also different from the one used in this work. 
Finally, regarding the MC runs, the intervals of randomly selected $\beta$ and $\gamma$ values are larger in this present work ($\pm 20 \times 10^{-5}$) compared to  \citep{2015CeMDA.123..325F} ($\pm 15 \times 10^{-5}$). Additionally, with the genetic algorithm method, one can not demonstrate that the  \citep{2015CeMDA.123..325F} convergency had reached the unique extremum (and not a local extremum), or even if such unique extremum does exist \citep{Katoch2021}. Using a GA approach is then problematic and has been not used for this work. Comparisons with the present results are then difficult.




\subsubsection{ $\dot{\mu}/\mu$}

\begin{figure}
    \centering
    \includegraphics[width=9.5cm,height=8cm]{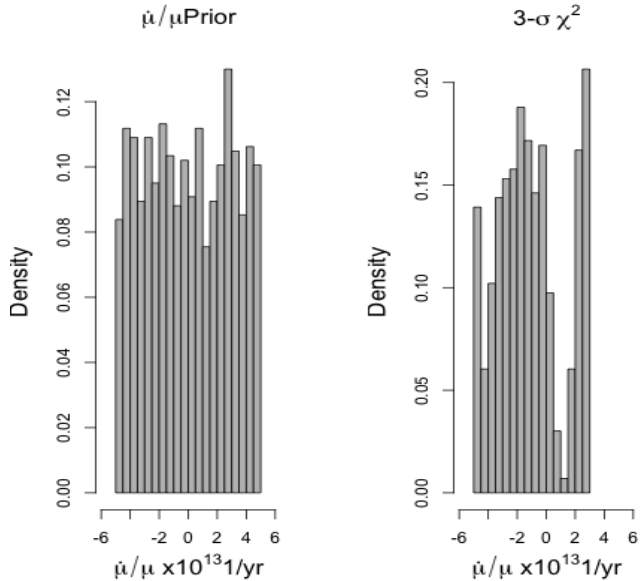}
        \caption{1D-histograms of $\dot{\mu}/\mu$ in yr$^{-1}$. The first left-hand side plot shows the uniform distribution of the prior when the middle histograms give the distribution of the selected values of $\eta$ according to the 3-$\sigma$ INPOP20a WRSS  distribution. }
    \label{fig:gdot1}
\end{figure}

The same procedure has been used for $\dot{\mu}/\mu$ using the INPOP20a datasets. 
Fig. \ref{fig:gdot1} gives the 1-D histograms of the initial distribution of $\dot{\mu}/\mu$ and the distribution of $\dot{\mu}/\mu$ corresponding to alternative ephemerides selected according to the 3-$\sigma$ WRSS  method.
A limit of about 2.3  $\times 10^{-13}$ yr$^{-1}$ is obtained with 60$\%$ of the runs selected.  If one supposes a constant Sun mass loss of about $(-0.92 \pm 0.61) \times 10^{-13}$ yr$^{-1}$, this result leads to $\dot{G}/G = (-0.08 \pm 2.84) \times 10^{-13}$ yr$^{-1}$, still consistent with GRT.
In terms of LS, $\dot{\mu}/\mu$ shows small correlations (less than 0.5) with the rest of the parameters involved in the INPOP20a adjustment.The LS covariance, $\pm 0.32 \times 10^{-13}$ yr$^{-1}$, is still smaller than the interval obtained with 3-$\sigma$ WRSS  filtering demonstrating again that this latest approach is more conservative then the direct LS.
Besides, these results are consistent with the   3-$\sigma$  LS uncertainties given by \citep{2015CeMDA.123..325F} but higher than their Monte Carlo estimates. As for $\beta$ and $\gamma$, this can be explained by the larger interval of random values explored in this work, $\pm 6 \times 10^{-13}$ yr$^{-1}$, in comparison with the interval used in \citep{2015CeMDA.123..325F},  $\pm 4 \times 10^{-13}$ yr$^{-1}$.

\subsection{Adding Bepi-Colombo simulations}
\label{sec:resbc}

\subsubsection{Planetary orbits and other fitted parameters}
\label{sec:cdi}

\begin{figure}
    \centering
    \includegraphics[width=9.5cm,height=9cm]{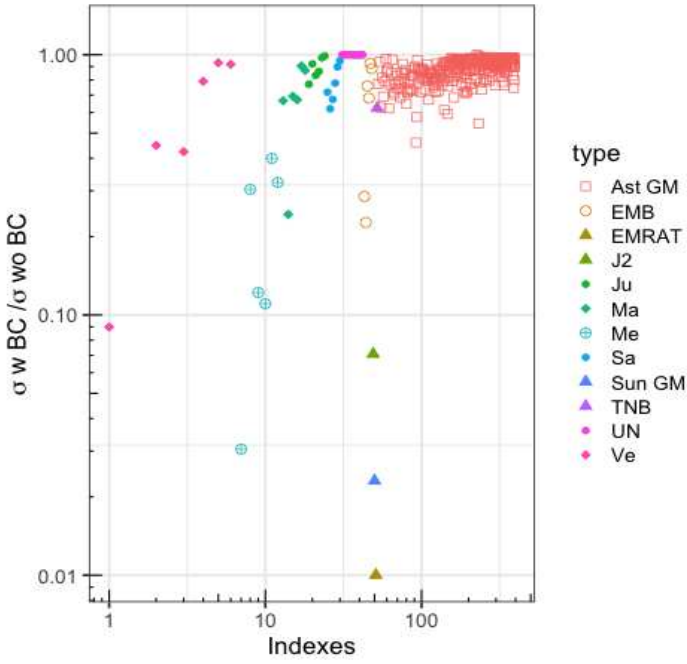}
    \caption{Distribution of the ratio between the parameter uncertainties obtained with and without BC simulations in log scale.The colors and shapes indicate the different types of parameters considered in the INPOP adjustment : Me,V,Ma,Ju,Sa,UN, EMB represent the ratio of the uncertainties for the 6 orbital initial conditions for Mercury, Venus, Mars, Jupiter, Saturn, Uranus, Neptune and the Earth-Moon barycenter respectively. $J^{\odot}_{2}$, Sun GM, EMRAT and TNB give the ratios for the Sun oblateness and mass, the ratio between the Earth and the Moon masses and the mass of the TNO ring respectively. Finally Ast GM indicate the ratio for the 343 Main Belt asteroid masses.}
    \label{fig:sigcdi}
\end{figure}

The first aspect to consider when one introduces BC simulations in planetary adjustments is the impact on the determination of planetary orbits by the means of the evolution of the covariance matrix of the planetary orbit initial conditions and other parameters of the fit.
On Fig. \ref{fig:sigcdi}, we plotted the ratio between the standard deviations (defined as the square root of the diagonal terms of the covariance matrix deduced from the least squares adjustment) for the 402 parameters of INPOP20a in GRT obtained in including the BC simulation ($\sigma$ w BC) and  without the BC simulation ($\sigma$ wo BC).
As one can see on this figure, the introduction of the BC simulation does not introduce any degradation of the parameter uncertainties as no ratio is greater than 1. The highly improved parameters, besides the Mercury and the Earth-Moon barycenter orbits,  are the Earth-Moon mass ratio, the mass of the Sun and its oblateness. The ratio between the variances obtained with and without BC are of at least of one order of magnitude for these parameters thanks to a better constraint on the Mercury geocentric orbit perturbed by the sun, provided by the BC observations. As secondary perturbers of the Mercury-Earth distance, Venus sees also its orbit improved as well as Mars. At a lower level, Jupiter and Saturn orbits are also better estimated when the other outer planet orbits are almost insensitive to the BC introduction. The determination of Main Belt asteroid masses does not seem to be drastically improved even if a noticeable increase of the ratio from 0.75 to 1 is visible. This indicates a slight reduction of the mass uncertainties for some of the perturbers.

\subsubsection{Sun core rotation}
\label{sec:J2bc}

In GRT, the results of the $J^{\odot}_{2}$ LS adjustment including BC simulations for different models of Sun core rotations are given in Table \ref{tab:SunBC}. 
One can notice a significant reduction of the 3-$\sigma$ LS uncertainty from 3$\times 10^{-9}$ with INPOP20a to 2 $\times 10^{-10}$ when including the BC simulations (as noticeable in Fig. \ref{fig:sigcdi} as well). At this level of accuracy, the differences between solar core rotation hypothesis appear to be detectable thanks to BC. 
More precisely, at the first glance, considering the four Sun core rotation modeles, no significant differences are noticeable in terms of WRSS  (Column 3 of Table \ref{tab:SunBC}) as they remain smaller than the interval of 3-$\sigma$ $\chi^{2}$, $\pm 0.03$, defined in Sect \ref{sec:method}. This means that these ephemerides are acceptable for the WRSS  filtering whatever the model for Sun core rotation. However, despite the fact that the estimated values for the Sun oblateness are still consistent at 2$\sigma$ with the \citep{2008A&A...477..657A} value, they differ from one Sun core rotation to another by a maximum of 5 $\times 10^{-10}$ (between the slow and the very fast rotations) , which is more than 2 times bigger than the 3-$\sigma$ LS uncertainty.  This could indicate a possible detection of the Sun core rotation thanks to the addition of the BC data.
In non-GRT (see Table \ref{tab:sres}), when we consider the 3-$\sigma$ WRSS  filtering , the addition of BC simulations induces that only 1 $\%$ of the computed runs reach the helioseismological limits. The interval of the fitted $J^{\odot}_{2}$ deduced from the selected WRSS  alternative ephemerides is about $\pm 1 \times 10^{-9}$.  This corresponds to an improvement of a factor 3 relative to the WRSS fitted $J^{\odot}_{2}$  obtained without BC, but it is  5 times larger than the LS 3-$\sigma$ uncertainty, $\pm 2.3 \times 10^{-10} $, obtained in Table \ref{tab:SunBC} by direct adjustment in GRT.  
If we consider the direct fit of $J^{\odot}_{2}$ together with the non-GRT parameters, the obtained 3-$\sigma$ uncertainty is improved relative to the fit without BC of about a factor almost 10, but it remains 5 times larger than the uncertainty obtained in GRT. 
In this context, the detection of the different models for the Sun core rotation appears then to be out of reach when we consider a simultaneous estimation of non-GRT parameters and $J^{\odot}_{2}$.
We can conclude that, there are  some indications of a possible detection of the Sun angular momentum from future BC observations in GRT with direct LS adjustment. 
Such detections seem to be difficult if  tests of non-GRT are done simultaneously, even when considering the WRSS  filtering. 

\begin{table}
\caption{Sun Angular Momentum and oblateness obtained in GRT (with PPN parameters fixed to unity) considering INPOP20a data samples and BC simulations. In Column 4, we give the differences  $\Delta J^{\odot}_{2}$ between the reference  $J^{\odot}_{2}$ fitted using \citep{1998MNRAS.297L..76P} angular momentum value and the  $J^{\odot}_{2}$ obtained after fit using different values for the amplitude of the angular momentum given in Column 2. Different models of Sun core rotation (identified in Column 1) are used for estimating S. The fourth  column gives the differences between the reference WRSS   and the WRSS  obtained for different Sun core rotations.  The uncertainties are given at 3-$\sigma$. See Sect \ref{sec:srot} for the significance of the different rotation modeling.}
\begin{tabular}{l c c c}
\hline
Type of rotation & S $\times 10^{48}$  & $\Delta \chi^{2}$ &  $\Delta J^{\odot}_{2} $\\
& g.cm$^{-2}$.s$^{-1}$& $\times 10^{4}$ &  $\times 10^{10}$\\ 
\hline
 INPOP20a + BC & 1.90 $\pm$ 1.5 &   0 &  0.0 $\pm$ 2.3\\
\\
Slow rotation &1.896 &  -4 &  -2 $\pm$ 2.3\\
Uniform at 435 nHz & 1.926  & -2 &  0.0 $\pm$ 2.3\\
Fast rotation&1.976 &  1 & 2  $\pm$ 2.3\\
Very fast rotation & 1.998 &  3 &  3 $\pm$ 2.3\\
\hline
\end{tabular}
\label{tab:SunBC}
\end{table}

\subsubsection{PPN parameters $\beta$, $\gamma$ and $\dot{\mu}/\mu$}
\label{sec:bg_bc}

\begin{figure}
    \centering
\includegraphics[scale=0.5]{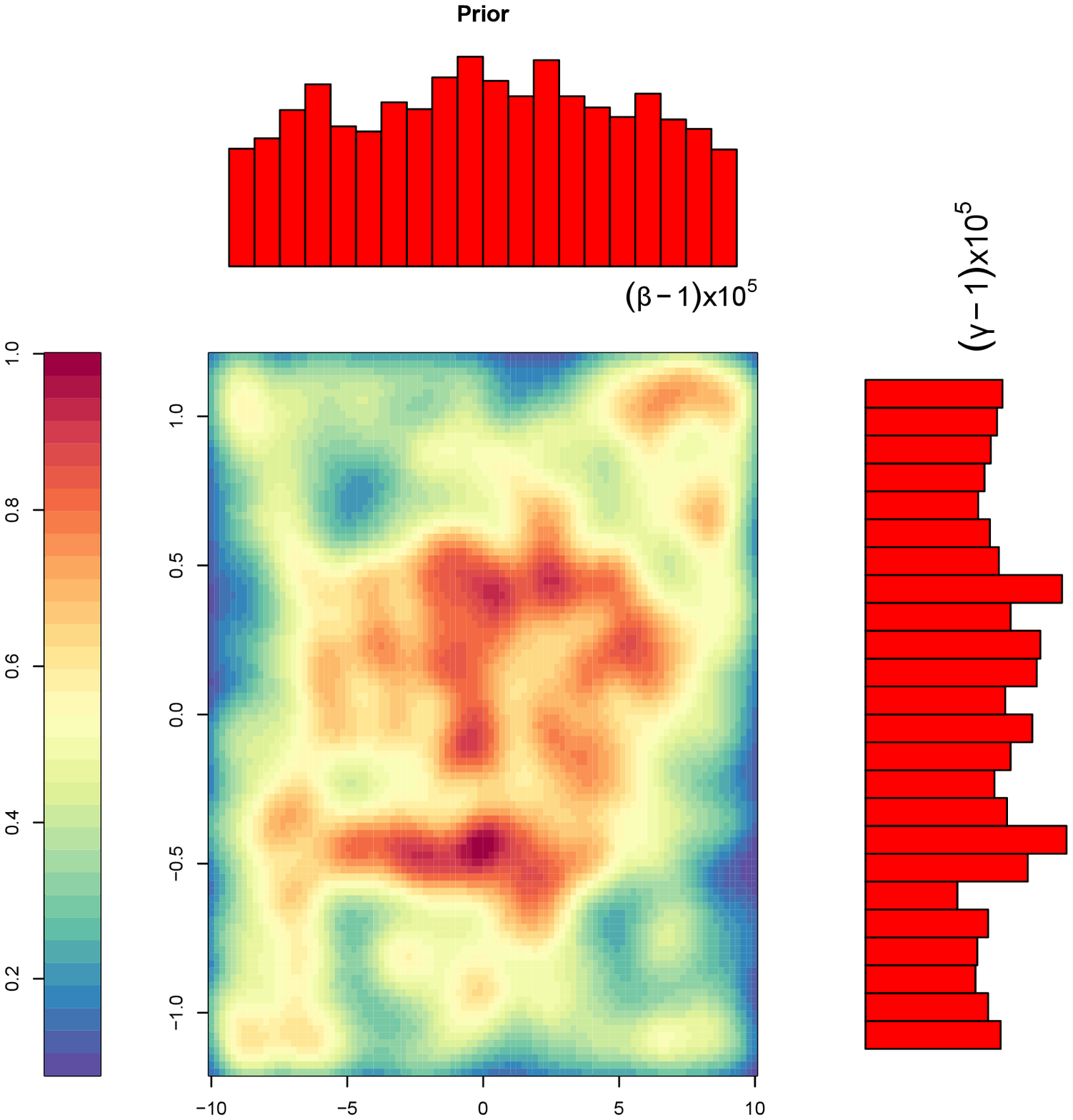}\includegraphics[scale=0.5]{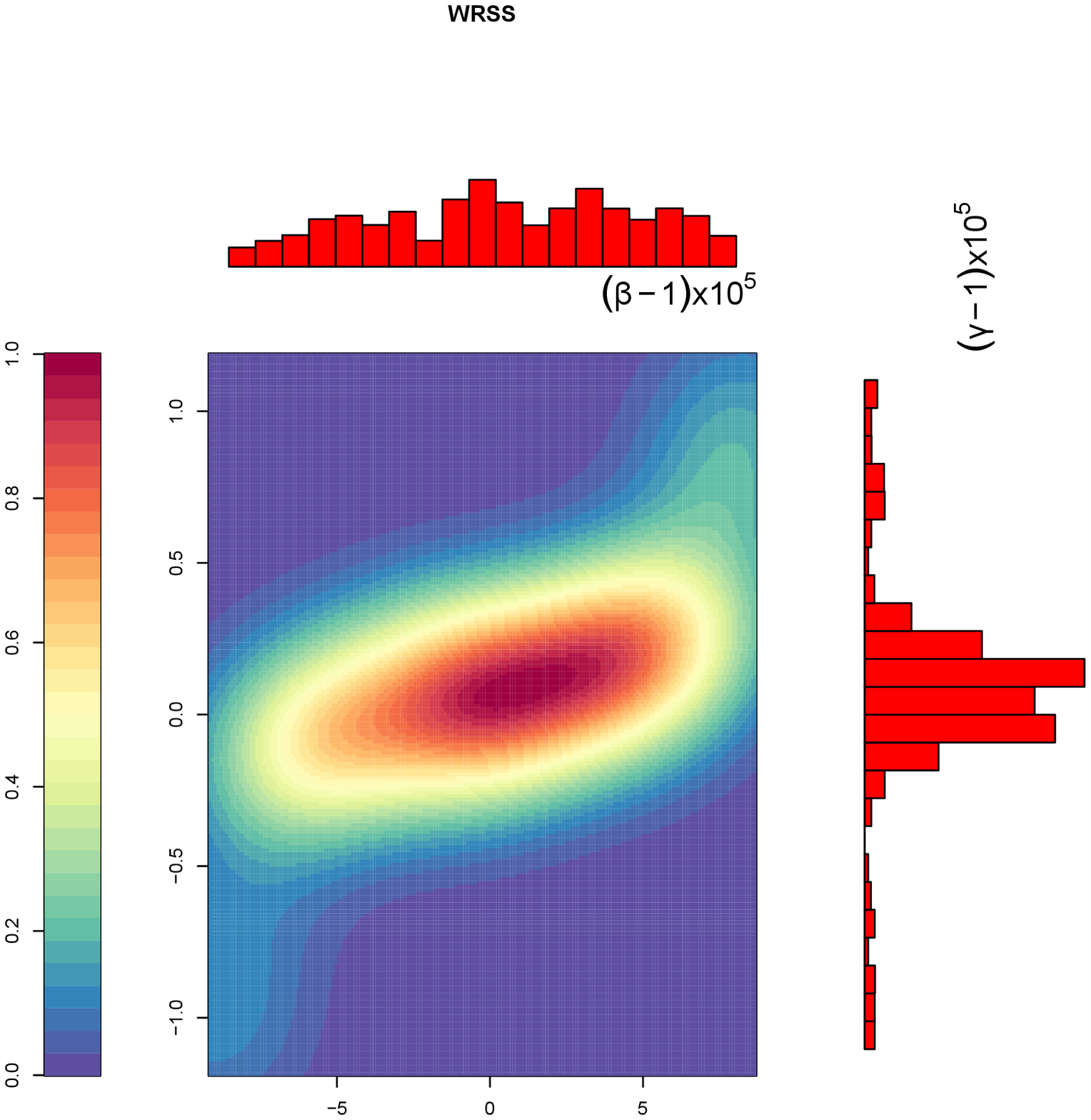}
        \caption{2D-histograms of PPN $\beta$ and $\gamma$ selected according to the 3-$\sigma$ INPOP20a WRSS  distribution in considering the Bepi-Colombo simulations. The color-scale indicates the normalised probability.}
    \label{fig:2DwBC}
\end{figure}

\begin{figure}
    \centering
    \includegraphics[width=9.5cm,height=8cm]{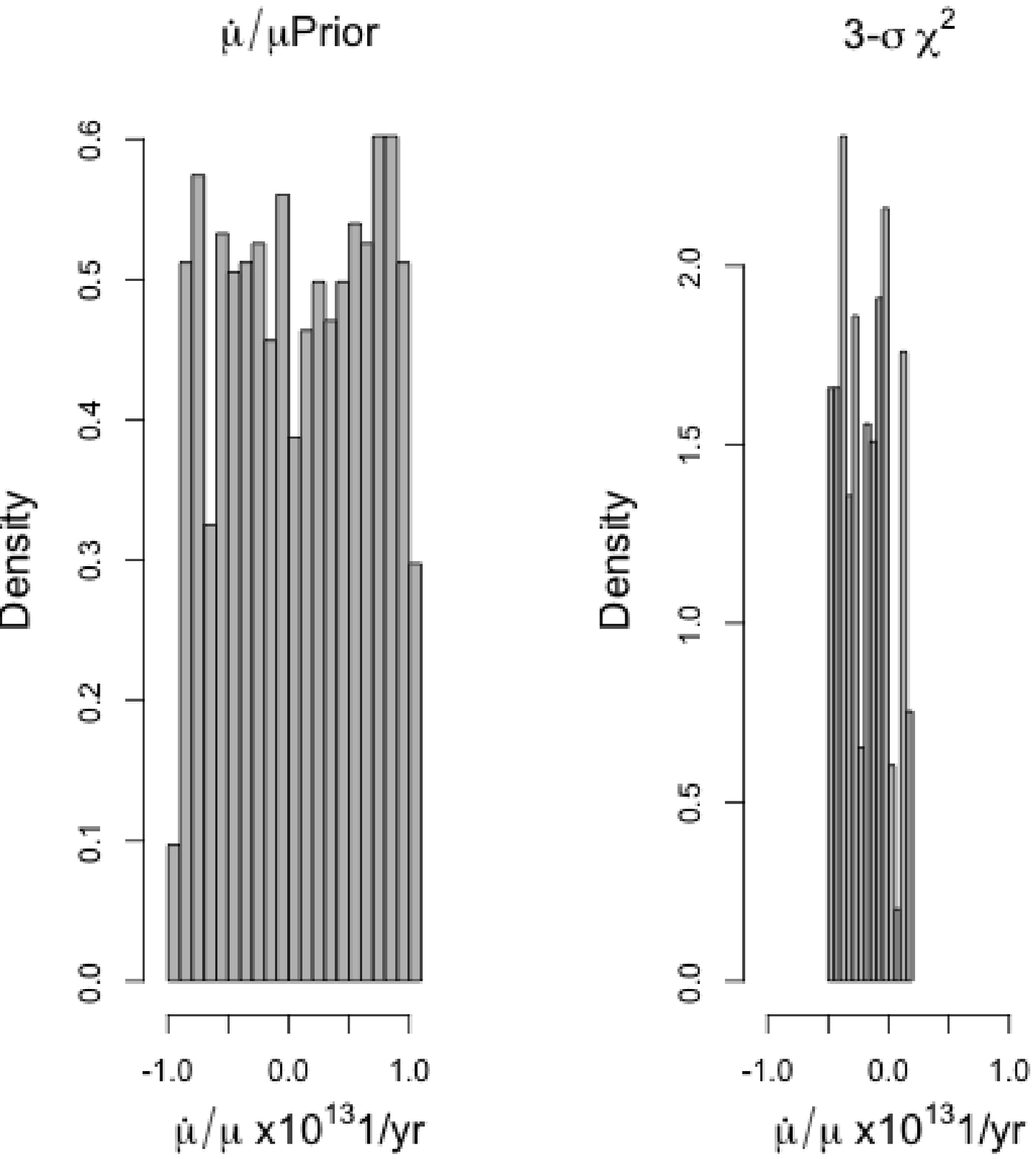}
        \caption{1D-histograms of $\dot{\mu}/\mu$ in yr$^{-1}$ including Bepi-Colombo simulations. The first left-hand side plot shows the uniform distribution of the prior when the middle histogramms give the distribution of the selected values of $\eta$ according to the 3-$\sigma$ INPOP20a WRSS  distribution considering the Bepi-Colombo simulations.}
    \label{fig:gdot2}
\end{figure}

On Table \ref{tab:sres} and Fig \ref{fig:2DwBC} are given the results obtained by adding the BC simulations to the INPOP20a data sample. A first striking result is  that BC will improve drastically the constraint on the possible violation of GRT through the PPN parameters $\beta$ and $\gamma$.
For the 3-$\sigma$ WRSS  filtering, the most spectacular is the estimation of the $\gamma$ parameter which gains a factor 19 in comparison with the INPOP20a results (see  also Fig. \ref{fig:2DwoBC}). The constraint on $\beta$ is less improved, of about a factor 1.5. We also note an improvement of the LS results with and without BC of about a factor 6 for $\beta$ and 15 for $\gamma$. 
These differences between $\beta$ and  $\gamma$ can again be explained by the introduction of the Lense-Thirring acceleration into the dynamical modeling of the planetary motion, that allows for a more efficient disentangling of the two parameters (see Sect. \ref{sec:resppn20a} and the correlation discussion). The introduction of the BC simulations also reduces the number of selected alternative ephemerides as only 8$\%$ of the alternative ephemerides have been selected with the 3-$\sigma$ WRSS  filtering.
On Fig. \ref{fig:gdot2}, are plotted the distributions of $\dot{\mu}/\mu$ before and after the 3-$\sigma$ WRSS  filtering including BC simulations.  27$\%$ of the runs are selected leading to a reduction of the interval of possible  $\dot{\mu}/\mu$ values of a factor 12. For the direct LS estimate, the improvement is even more important, of about a factor 30.
With such a constraint, in the perspective of measuring $\dot{G}/G$, it will be important to have independent and accurate constraints for the Sun mass loss which has currently a higher uncertainty ($0.61 \times 10^{-13}$ yr$^{-1}$). 

\label{sec:mu_bc}

\section{Discussion and conclusion}
\label{sec:final}

\subsection{Helioseismology and variations of the Sun mass}

\begin{figure}
    \centering
    \includegraphics[scale=0.5]{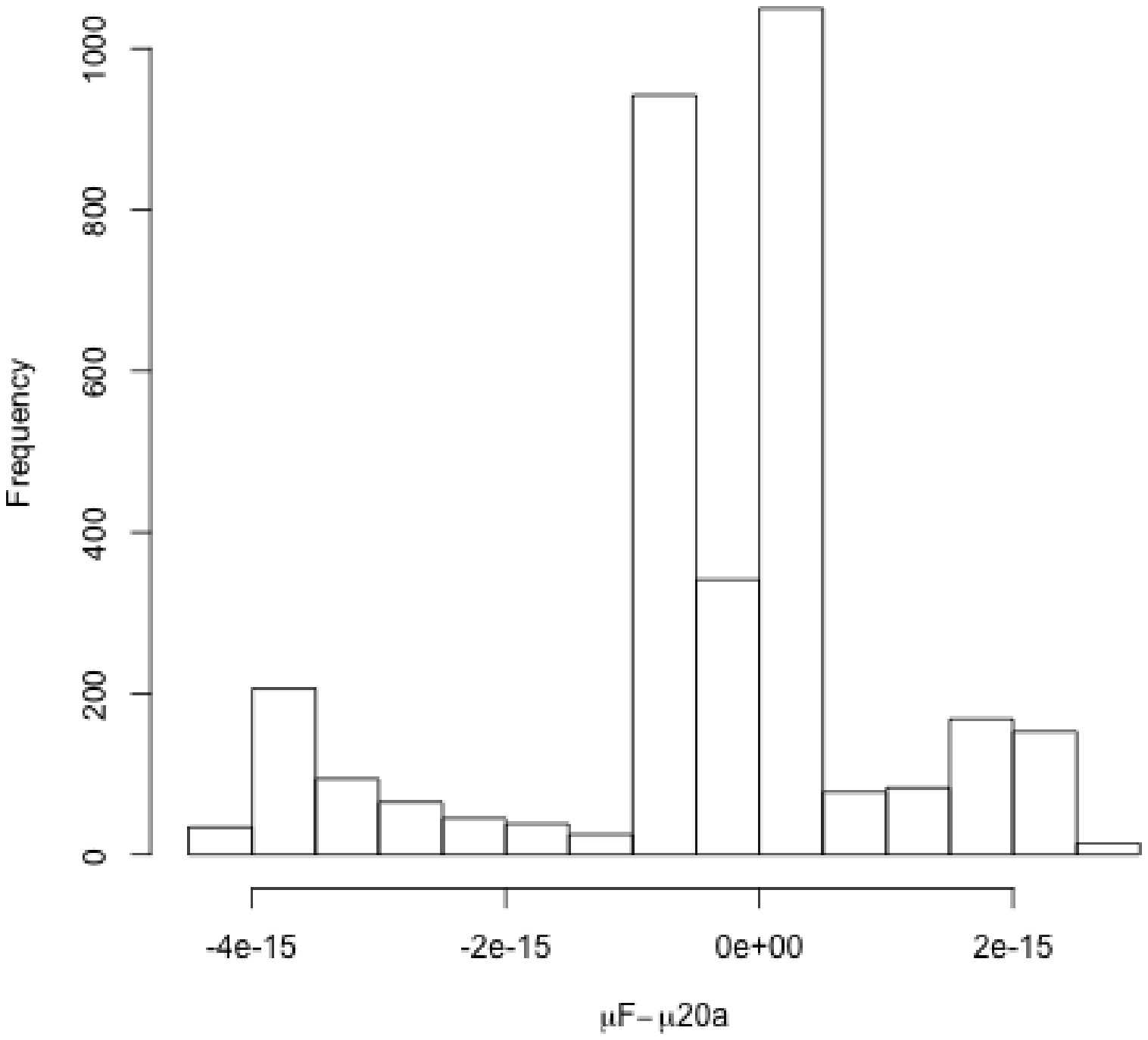}
    \caption{Differences in UA$^{3}$.d$^{-2}$ between INPOP20a Sun $\mu$ and $\mu$ fitted over 4500 alternative ephemerides built with  non-unity PPN $\beta$ and $\gamma$ and with non-zero $\eta$ and $\dot{\mu}/\mu$.}
    \label{fig:mu}
\end{figure}

Limits for the possible variations of non-GRT parameters ($\beta$, $\gamma$ and $\dot{\mu}/\mu$) have been obtained with Sun oblateness thresholds based on helioseismology measurements (see Sect \ref{sec:LT}). These measures were obtained in considering the variations of the Sun angular momentum and its kinematic energy for a fixed value of the gravitational mass of the Sun, $\mu$. So it is interesting to address the question of how could change the value of the helioseismological  Sun oblateness for different values of $\mu$. From \citep{1981MNRAS.196..731G,1998MNRAS.297L..76P} we see that the $J^{\odot}_{2}$ measurement deduced from helioseismology relies on the following equation 
$$
J^{\odot}_{2} = - \frac{R_{\odot}}{\mu} \phi_{12}(R_{\odot}),
$$
where $R_{\odot}$ is the radius of the Sun, $\mu$ its gravitational mass and $\phi_{12}$ is the quadrupole component of the gravitational potential, deduced from the Poisson equation.  $\phi_{12}$ depends on $R_{\odot}$ and consequently, we can estimate the impact on the estimation of $J^{\odot}_{2}$, $\delta (J^{\odot}_{2})$, of introducing a change in the gravitational mass $\delta \mu$ for a fixed value of $R_{\odot}$  by 
\begin{equation}
\centering
\frac{\delta (J^{\odot}_{2})}{J^{\odot}_{2}} = - \frac{1}{2} \frac{\delta \mu}{\mu}.
\label{eq:muSun}
\end{equation}
In Sec \ref{sec:LT}, we consider possible variations of $J^{\odot}_{2}$ into the range of   $ \pm 1.5 \times 10^{-8}$. For such an interval of $\delta (J^{\odot}_{2})$ the equation \ref{eq:muSun} gives a corresponding variation of $\mu$, $\delta \mu$, of about  $4 \times 10^{-5}$ UA$^{3}$.d$^{-2}$. This means that for inducing a change in the helioseismic estimation of $J^{\odot}_{2}$ greater than the interval considered in Sec \ref{sec:LT}, we need to introduce variations of $\mu$ of the order of $10^{-5}$ UA$^{3}$.d$^{-2}$.
However, as one can see on the histogram of Fig. \ref{fig:mu}, the distribution of the differences between the INPOP20a $\mu$ and the values estimated for the alternative ephemerides are clearly under this threshold, the maximum difference being of about $4 \times 10^{-15}$ UA$^{3}$.d$^{-2}$. With such a difference, the impact on the helioseismic $J^{\odot}_{2}$ is of about $10^{-18}$.
We can also note that the differences between 1998-published DE405 gravitational mass \citep{Standish2001} and INPOP20a is about 3$\times 10^{-11}$ UA$^{3}$.d$^{-2}$, again below  $10^{-5}$ UA$^{3}$.d$^{-2}$.
We can then conclude that even if we consider different values of $\mu$ compared to the value used by \citep{1998MNRAS.297L..76P}, the impact on the $J^{\odot}_{2}$ determinations is clearly encompassed in the interval of uncertainty used in this work.

\subsection{Comparisons with previous results}

In 2015, a similar approach based on Monte Carlo sampling of PPN parameters, was proposed by \citep{2015CeMDA.123..325F} using INPOP15a. Since then, improvements have been brought to the adjustment method, especially regarding asteroid mass determination \citep{2020MNRAS.492..589F}, and  the addition of accurate data for Mercury, Mars, Jupiter and Saturn. Values obtained by the \citep{2015CeMDA.123..325F} Monte Carlo sampling are presented for comparison in Table \ref{tab:sres}. The intervals obtained with INPOP20a show a clear improvement for  parameters  estimated by \citep{2015CeMDA.123..325F}. 
In 2020, \citep{2020CQGra..37i5007D} presented a covariance analysis of least square determinations of the same non-GRT parameters. They consider different sets of tracking data used for planetary ephemerides construction (Messenger, Mars orbiters, Juno and Cassini missions) as well as future missions including BC. 
Before comparing our results with theirs, it is important to note that  \citep{2020CQGra..37i5007D} introduce a linear relation (Nordvedt relation) between the PPN parameters $\beta$, $\gamma$ and $\eta$ as $\eta= 4(\beta-1) - (\gamma -1) - \alpha_{1} - \frac{2}{3} \alpha_{2}$
where $\eta$ is the ratio between inertial mass and gravitational mass, accounting for the strong equivalence test and $\alpha_{1}$ and $\alpha_{2}$ testing preferred-frame hypothesis, can be consider as 0.
By introducing such a relation, it has been demonstrated that the uncertainty on $\beta$ estimations is severely diminished (see for example \citep{2018Icar..301....9I}) but at the cost of generality in terms of possible theories to be tested. This is for this reason that we do not introduce this relation in our work.
With the present INPOP20a data sets, our LS results are in good agreement with  \citep{2020CQGra..37i5007D} covariances. In comparison, the intervals produced by the 3-$\sigma$ WRSS  filtering are larger for all parameters, showing again that this method is more conservative relative to the LS estimation or the covariance analysis.
The improvement brought by BC simulations on the 3-$\sigma$ WRSS  determinations of the non-GRT parameters are very close to the improvements proposed by \citep{2020CQGra..37i5007D}. We find that the ratio between the  3-$\sigma$ WRSS  acceptable intervals with and without BC for $\gamma-1$, Sun $J^{\odot}_{2}$,  $\beta-1$ are quite close or of the same order of magnitude than  \citep{2020CQGra..37i5007D} despite the fact that we did not include the Nordvedt relation. For $\dot{\mu}/\mu$, however,  we note an improvement induced by BC simulations of about a factor 10 when \citep{2020CQGra..37i5007D} indicate a factor 5.

\subsection{Conclusion}

In this study, we give a detailed description of the latest INPOP planetary ephemerides INPOP20a as well as new evaluations of possible GRT violations with the PPN parameters $\beta$,$\gamma$ and  $\dot{\mu}/\mu$. With a new method for selecting acceptable alternative ephemerides we provide conservative limits of about $7.16 \times 10^{-5}$ and  $7.49 \times 10^{-5}$ for $\beta-1$ and $\gamma-1$ respectively using the present day planetary data samples. We show, as already stated in \citep{2018Icar..301....9I, 2020CQGra..37i5007D}, that the use of future BC range observations should improve these estimates, in particular $\gamma$. 
Limits of possible secular variations of the Sun gravitational mass are given with a limit of about $2.28 \times 10^{-13}$ $yr^{-1}$ without BC simulations and  $0.19 \times 10^{-13}$ $yr^{-1}$ with. Finally, interesting perspectives for the detection of the Sun core rotation seem to be reachable thanks to the BC mission and its accurate range measurements in the GRT frame.

 \bibliography{biblio_asteroid_mass}

\begin{thebibliography}{44}
\providecommand{\natexlab}[1]{#1}
\providecommand{\url}[1]{\texttt{#1}}
\expandafter\ifx\csname urlstyle\endcsname\relax
  \providecommand{\doi}[1]{doi: #1}\else
  \providecommand{\doi}{doi: \begingroup \urlstyle{rm}\Url}\fi

\bibitem[{Antia} et~al.(2008){Antia}, {Chitre}, and
  {Gough}]{2008A&A...477..657A}
H.~M. {Antia}, S.~M. {Chitre}, and D.~O. {Gough}.
\newblock {Temporal variations in the Sun's rotational kinetic energy}.
\newblock \emph{\aap}, 477\penalty0 (2):\penalty0 657--663, January 2008.
\newblock \doi{10.1051/0004-6361:20078209}.

\bibitem[{Appourchaux} and {Corbard}(2019)]{2019A&A...624A.106A}
T.~{Appourchaux} and T.~{Corbard}.
\newblock {Searching for g modes. II. Unconfirmed g-mode detection in the power
  spectrum of the time series of round-trip travel time}.
\newblock \emph{\aap}, 624:\penalty0 A106, April 2019.
\newblock \doi{10.1051/0004-6361/201935196}.

\bibitem[{Archinal} et~al.(2018){Archinal}, {Acton}, {A'Hearn}, {Conrad},
  {Consolmagno}, {Duxbury}, {Hestroffer}, {Hilton}, {Kirk}, {Klioner},
  {McCarthy}, {Meech}, {Oberst}, {Ping}, {Seidelmann}, {Tholen}, {Thomas}, and
  {Williams}]{2018CeMDA.130...22A}
B.~A. {Archinal}, C.~H. {Acton}, M.~F. {A'Hearn}, A.~{Conrad}, G.~J.
  {Consolmagno}, T.~{Duxbury}, D.~{Hestroffer}, J.~L. {Hilton}, R.~L. {Kirk},
  S.~A. {Klioner}, D.~{McCarthy}, K.~{Meech}, J.~{Oberst}, J.~{Ping}, P.~K.
  {Seidelmann}, D.~J. {Tholen}, P.~C. {Thomas}, and I.~P. {Williams}.
\newblock {Report of the IAU Working Group on Cartographic Coordinates and
  Rotational Elements: 2015}.
\newblock \emph{Celestial Mechanics and Dynamical Astronomy}, 130\penalty0
  (3):\penalty0 22, February 2018.
\newblock \doi{10.1007/s10569-017-9805-5}.

\bibitem[{Bernus}(2020)]{Bernus2020b}
L.~{Bernus}.
\newblock \emph{Tests de graviation \`a l'\'echelle du systeme solaire}.
\newblock PhD thesis, Observatoire de Paris, 2020.

\bibitem[{Bernus} et~al.(2019){Bernus}, {Minazzoli}, {Fienga}, {Gastineau},
  {Laskar}, and {Deram}]{bernus2019}
L.~{Bernus}, O.~{Minazzoli}, A.~{Fienga}, M.~{Gastineau}, J.~{Laskar}, and
  P.~{Deram}.
\newblock {Constraining the Mass of the Graviton with the Planetary Ephemeris
  INPOP}.
\newblock \emph{\prl}, 123\penalty0 (16):\penalty0 161103, October 2019.
\newblock \doi{10.1103/PhysRevLett.123.161103}.

\bibitem[{Bernus} et~al.(2020){Bernus}, {Minazzoli}, { Fienga}, {Gastineau},
  {Laskar}, {Deram}, and {Di Ruscio}]{2020PhRvD.102b1501B}
L.~{Bernus}, O.~{Minazzoli}, A.~{ Fienga}, M.~{Gastineau}, J.~{Laskar},
  P.~{Deram}, and A.~{Di Ruscio}.
\newblock {Constraint on the Yukawa suppression of the Newtonian potential from
  the planetary ephemeris INPOP19a}.
\newblock \emph{\prd}, 102\penalty0 (2):\penalty0 021501, July 2020.
\newblock \doi{10.1103/PhysRevD.102.021501}.

\bibitem[{Bertotti} et~al.(2003){Bertotti}, {Iess}, and
  {Tortora}]{bertotti2003nat}
B.~{Bertotti}, L.~{Iess}, and P.~{Tortora}.
\newblock {A test of general relativity using radio links with the Cassini
  spacecraft}.
\newblock \emph{\nat}, 425\penalty0 (6956):\penalty0 374--376, September 2003.
\newblock \doi{10.1038/nature01997}.

\bibitem[{Chaplin} et~al.(1999){Chaplin}, {Christensen-Dalsgaard}, {Elsworth},
  {Howe}, {Isaak}, {Larsen}, {New}, {Schou}, {Thompson}, and
  {Tomczyk}]{1999MNRAS.308..405C}
W.~J. {Chaplin}, J.~{Christensen-Dalsgaard}, Y.~{Elsworth}, R.~{Howe}, G.~R.
  {Isaak}, R.~M. {Larsen}, R.~{New}, J.~{Schou}, M.~J. {Thompson}, and
  S.~{Tomczyk}.
\newblock {Rotation of the solar core from BiSON and LOWL frequency
  observations}.
\newblock \emph{\mnras}, 308\penalty0 (2):\penalty0 405--414, September 1999.
\newblock \doi{10.1046/j.1365-8711.1999.02691.x}.

\bibitem[{Christensen-Dalsgaard}(2021)]{Christensen-Dalsgaard2021}
J.~{Christensen-Dalsgaard}.
\newblock Private communication, 2021.

\bibitem[{De Marchi} and {Cascioli}(2020)]{2020CQGra..37i5007D}
F.~{De Marchi} and G.~{Cascioli}.
\newblock {Testing general relativity in the solar system: present and future
  perspectives}.
\newblock \emph{Classical and Quantum Gravity}, 37\penalty0 (9):\penalty0
  095007, May 2020.
\newblock \doi{10.1088/1361-6382/ab6ae0}.

\bibitem[{Di Mauro}(2003)]{2003LNP...599...31D}
M.~P. {Di Mauro}.
\newblock \emph{{Helioseismology: A Fantastic Tool to Probe the Interior of the
  Sun}}, volume 599, pages 31--67.
\newblock 2003.

\bibitem[{Di Ruscio} et~al.(2020){Di Ruscio}, { Fienga}, {Durante}, {Iess},
  {Laskar}, and {Gastineau}]{2020A&A...640A...7D}
A.~{Di Ruscio}, A.~{ Fienga}, D.~{Durante}, L.~{Iess}, J.~{Laskar}, and
  M.~{Gastineau}.
\newblock {Analysis of Cassini radio tracking data for the construction of
  INPOP19a: A new estimate of the Kuiper belt mass}.
\newblock \emph{\aap}, 640:\penalty0 A7, August 2020.
\newblock \doi{10.1051/0004-6361/202037920}.

\bibitem[{Fienga} et~al.(2009){Fienga}, {Laskar}, {Morley}, {Manche},
  {Kuchynka}, {Le Poncin-Lafitte}, {Budnik}, {Gastineau}, and
  {Somenzi}]{2009A&A...507.1675F}
A.~{Fienga}, J.~{Laskar}, T.~{Morley}, H.~{Manche}, P.~{Kuchynka}, C.~{Le
  Poncin-Lafitte}, F.~{Budnik}, M.~{Gastineau}, and L.~{Somenzi}.
\newblock {INPOP08, a 4-D planetary ephemeris: from asteroid and time-scale
  computations to ESA Mars Express and Venus Express contributions}.
\newblock \emph{\aap}, 507:\penalty0 1675--1686, December 2009.
\newblock \doi{10.1051/0004-6361/200911755}.

\bibitem[{Fienga} et~al.(2015){Fienga}, {Laskar}, {Exertier}, {Manche}, and
  {Gastineau}]{2015CeMDA.123..325F}
A.~{Fienga}, J.~{Laskar}, P.~{Exertier}, H.~{Manche}, and M.~{Gastineau}.
\newblock {Numerical estimation of the sensitivity of INPOP planetary
  ephemerides to general relativity parameters}.
\newblock \emph{Celestial Mechanics and Dynamical Astronomy}, 123:\penalty0
  325--349, November 2015.
\newblock \doi{10.1007/s10569-015-9639-y}.

\bibitem[{Fienga} et~al.(2019){Fienga}, {Deram}, {Viswanathan}, {Di Ruscio},
  {Bernus}, {Durante}, {Gastineau}, and {Laskar}]{2019NSTIM.109.....V}
A.~{Fienga}, P.~{Deram}, V.~{Viswanathan}, A.~{Di Ruscio}, L.~{Bernus},
  D.~{Durante}, M.~{Gastineau}, and J.~{Laskar}.
\newblock {INPOP19a planetary ephemerides}.
\newblock \emph{Notes Scientifiques et Techniques de l'Institut de Mecanique
  Celeste}, 109, December 2019.

\bibitem[{Fienga} et~al.(2020){Fienga}, {Avdellidou}, and
  {Hanu{\v{s}}}]{2020MNRAS.492..589F}
A.~{Fienga}, C.~{Avdellidou}, and J.~{Hanu{\v{s}}}.
\newblock {Asteroid masses obtained with INPOP planetary ephemerides}.
\newblock \emph{\mnras}, 492\penalty0 (1):\penalty0 589--602, February 2020.
\newblock \doi{10.1093/mnras/stz3407}.

\bibitem[{Fossat} and {Schmider}(2018)]{2018A&A...612L...1F}
E.~{Fossat} and F.~X. {Schmider}.
\newblock {More about solar g modes}.
\newblock \emph{\aap}, 612:\penalty0 L1, April 2018.
\newblock \doi{10.1051/0004-6361/201832626}.

\bibitem[{Fossat} et~al.(2017){Fossat}, {Boumier}, {Corbard}, {Provost},
  {Salabert}, {Schmider}, {Gabriel}, {Grec}, {Renaud}, {Robillot},
  {Roca-Cort{\'e}s}, {Turck-Chi{\`e}ze}, {Ulrich}, and
  {Lazrek}]{2017A&A...604A..40F}
E.~{Fossat}, P.~{Boumier}, T.~{Corbard}, J.~{Provost}, D.~{Salabert}, F.~X.
  {Schmider}, A.~H. {Gabriel}, G.~{Grec}, C.~{Renaud}, J.~M. {Robillot},
  T.~{Roca-Cort{\'e}s}, S.~{Turck-Chi{\`e}ze}, R.~K. {Ulrich}, and M.~{Lazrek}.
\newblock {Asymptotic g modes: Evidence for a rapid rotation of the solar
  core}.
\newblock \emph{\aap}, 604:\penalty0 A40, August 2017.
\newblock \doi{10.1051/0004-6361/201730460}.

\bibitem[{Garc{\'\i}a} et~al.(2004){Garc{\'\i}a}, {Corbard}, {Chaplin},
  {Couvidat}, {Eff-Darwich}, {Jim{\'e}nez-Reyes}, {Korzennik}, {Ballot},
  {Boumier}, {Fossat}, {Henney}, {Howe}, {Lazrek}, {Lochard}, {Pall{\'e}}, and
  {Turck-Chi{\`e}ze}]{2004SoPh..220..269G}
R.~A. {Garc{\'\i}a}, T.~{Corbard}, W.~J. {Chaplin}, S.~{Couvidat},
  A.~{Eff-Darwich}, S.~J. {Jim{\'e}nez-Reyes}, S.~G. {Korzennik}, J.~{Ballot},
  P.~{Boumier}, E.~{Fossat}, C.~J. {Henney}, R.~{Howe}, M.~{Lazrek},
  J.~{Lochard}, P.~L. {Pall{\'e}}, and S.~{Turck-Chi{\`e}ze}.
\newblock {About the rotation of the solar radiative interior}.
\newblock \emph{\solphys}, 220\penalty0 (2):\penalty0 269--285, April 2004.
\newblock \doi{10.1023/B:SOLA.0000031395.90891.ce}.

\bibitem[{Garc{\'\i}a} et~al.(2007){Garc{\'\i}a}, {Turck-Chi{\`e}ze},
  {Jim{\'e}nez-Reyes}, {Ballot}, {Pall{\'e}}, {Eff-Darwich}, {Mathur}, and
  {Provost}]{2007Sci...316.1591G}
R.~A. {Garc{\'\i}a}, S.~{Turck-Chi{\`e}ze}, S.~J. {Jim{\'e}nez-Reyes},
  J.~{Ballot}, P.~L. {Pall{\'e}}, A.~{Eff-Darwich}, S.~{Mathur}, and
  J.~{Provost}.
\newblock {Tracking Solar Gravity Modes: The Dynamics of the Solar Core}.
\newblock \emph{Science}, 316\penalty0 (5831):\penalty0 1591, June 2007.
\newblock \doi{10.1126/science.1140598}.

\bibitem[{Gavryuseva} et~al.(1998){Gavryuseva}, {Gavryusev}, and {Di
  Mauro}]{1998ESASP.418..193G}
E.~{Gavryuseva}, V.~{Gavryusev}, and M.~P. {Di Mauro}.
\newblock {Rotational Split of Solar Acoustic Modes from GONG Experiment}.
\newblock In S.~{Korzennik}, editor, \emph{Structure and Dynamics of the
  Interior of the Sun and Sun-like Stars}, volume 418 of \emph{ESA Special
  Publication}, page 193, January 1998.

\bibitem[{Genova} et~al.(2018){Genova}, {Mazarico}, {Goossens}, {Lemoine},
  {Neumann}, {Smith}, and {Zuber}]{2018NatCo...9..289G}
A.~{Genova}, E.~{Mazarico}, S.~{Goossens}, F.G. {Lemoine}, G.~A. {Neumann},
  D.~E. {Smith}, and M.~T. {Zuber}.
\newblock {Solar system expansion and strong equivalence principle as seen by
  the NASA MESSENGER mission}.
\newblock \emph{Nature Communications}, 9:\penalty0 289, January 2018.
\newblock \doi{10.1038/s41467-017-02558-1}.

\bibitem[{Gough}(1981)]{1981MNRAS.196..731G}
D.~O. {Gough}.
\newblock {A new measure of the solar rotation}.
\newblock \emph{\mnras}, 196:\penalty0 731--745, September 1981.
\newblock \doi{10.1093/mnras/196.3.731}.

\bibitem[{Gough}(2015)]{2015SSRv..196...15G}
D.~O. {Gough}.
\newblock {Some Glimpses from Helioseismology at the Dynamics of the Deep Solar
  Interior}.
\newblock \emph{\ssr}, 196\penalty0 (1-4):\penalty0 15--47, December 2015.
\newblock \doi{10.1007/s11214-015-0159-6}.

\bibitem[{Hees}(2015)]{Hees2015}
A.~{Hees}.
\newblock Private communication, 2015.

\bibitem[{Iess} et~al.(2021){Iess}, {Asmar}, {Cappuccio}, {Cascioli}, {De
  Marchi}, {di Stefano}, {Genova}, {Ashby}, {Barriot}, {Bender}, {Benedetto},
  {Border}, {Budnik}, {Ciarcia}, {Damour}, {Dehant}, {Di Achille}, {Di Ruscio},
  {Fienga}, {Formaro}, {Klioner}, {Konopliv}, {Lema{\^\i}tre}, {Longo},
  {Mercolino}, {Mitri}, {Notaro}, {Olivieri}, {Paik}, {Palli}, {Schettino},
  {Serra}, {Simone}, {Tommei}, {Tortora}, {Van Hoolst}, {Vokrouhlick{\'y}},
  {Watkins}, {Wu}, and {Zannoni}]{2021SSRv..217...21I}
L.~{Iess}, S.~W. {Asmar}, P.~{Cappuccio}, G.~{Cascioli}, F.~{De Marchi}, I.~{di
  Stefano}, A.~{Genova}, N.~{Ashby}, J.~P. {Barriot}, P.~{Bender},
  C.~{Benedetto}, J.~S. {Border}, F.~{Budnik}, S.~{Ciarcia}, T.~{Damour},
  V.~{Dehant}, G.~{Di Achille}, A.~{Di Ruscio}, A.~{Fienga}, R.~{Formaro},
  S.~{Klioner}, A.~{Konopliv}, A.~{Lema{\^\i}tre}, F.~{Longo}, M.~{Mercolino},
  G.~{Mitri}, V.~{Notaro}, A.~{Olivieri}, M.~{Paik}, A.~{Palli},
  G.~{Schettino}, D.~{Serra}, L.~{Simone}, G.~{Tommei}, P.~{Tortora}, T.~{Van
  Hoolst}, D.~{Vokrouhlick{\'y}}, M.~{Watkins}, X.~{Wu}, and M.~{Zannoni}.
\newblock {Gravity, Geodesy and Fundamental Physics with BepiColombo's MORE
  Investigation}.
\newblock \emph{\ssr}, 217\penalty0 (1):\penalty0 21, February 2021.
\newblock \doi{10.1007/s11214-021-00800-3}.

\bibitem[{Imperi} and {Iess}(2017)]{2017CQGra..34g5002I}
L.~{Imperi} and L.~{Iess}.
\newblock {The determination of the post-Newtonian parameter
  {\ensuremath{\gamma}} during the cruise phase of BepiColombo}.
\newblock \emph{Classical and Quantum Gravity}, 34\penalty0 (7):\penalty0
  075002, April 2017.
\newblock \doi{10.1088/1361-6382/aa606d}.

\bibitem[{Imperi} et~al.(2018){Imperi}, {Iess}, and
  {Mariani}]{2018Icar..301....9I}
L.~{Imperi}, L.~{Iess}, and M.~J. {Mariani}.
\newblock {An analysis of the geodesy and relativity experiments of
  BepiColombo}.
\newblock \emph{Icarus}, 301:\penalty0 9025, February 2018.
\newblock \doi{10.1016/j.icarus.2017.09.008}.

\bibitem[{Katoch} and {Chauhan}(2021)]{Katoch2021}
S.~{Katoch} and S.~.S. {Chauhan}.
\newblock {A review on genetic algorithm: past, present, and future}.
\newblock \emph{Multimedia Tools and Applications}, 80\penalty0 (5):\penalty0
  8091--8126, February 2021.
\newblock \doi{10.1007/s11042-020-10139-6}.

\bibitem[{Komm} et~al.(2003){Komm}, {Howe}, {Durney}, and
  {Hill}]{2003ApJ...586..650K}
R.~{Komm}, R.~{Howe}, B.~R. {Durney}, and F.~{Hill}.
\newblock {Temporal Variation of Angular Momentum in the Solar Convection
  Zone}.
\newblock \emph{\apj}, 586\penalty0 (1):\penalty0 650--662, March 2003.
\newblock \doi{10.1086/367608}.

\bibitem[{Lazrek} et~al.(1996){Lazrek}, {Pantel}, {Fossat}, {Gelly},
  {Schmider}, {Fierry-Fraillon}, {Grec}, {Loudagh}, {Ehgamberdiev}, {Khamitov},
  {Hoeksema}, {Pall{\'e}}, and {R{\'e}gulo}]{1996SoPh..166....1L}
M.~{Lazrek}, A.~{Pantel}, E.~{Fossat}, B.~{Gelly}, F.~X. {Schmider},
  D.~{Fierry-Fraillon}, G.~{Grec}, S.~{Loudagh}, S.~{Ehgamberdiev},
  I.~{Khamitov}, J.~T. {Hoeksema}, P.~L. {Pall{\'e}}, and C.~{R{\'e}gulo}.
\newblock {Is the Solar Core Rotating Faster of Slower Than the Envelope?}
\newblock \emph{\solphys}, 166\penalty0 (1):\penalty0 1--16, June 1996.
\newblock \doi{10.1007/BF00179353}.

\bibitem[{Lense} and {Thirring}(1918)]{1918PhyZ...19..156L}
J.~{Lense} and H.~{Thirring}.
\newblock {{\"U}ber den Einflu{\ss} der Eigenrotation der Zentralk{\"o}rper auf
  die Bewegung der Planeten und Monde nach der Einsteinschen
  Gravitationstheorie}.
\newblock \emph{Physikalische Zeitschrift}, 19:\penalty0 156, January 1918.

\bibitem[{Moskovitz} et~al.(2018){Moskovitz}, {Schottland}, {Burt}, {Bailen},
  and {Wasserman}]{2018DPS....5040808M}
N.~{Moskovitz}, R.~{Schottland}, B.~{Burt}, M.~{Bailen}, and L.~{Wasserman}.
\newblock {astorb at Lowell Observatory: A comprehensive system to enable
  asteroid science}.
\newblock In \emph{AAS/Division for Planetary Sciences Meeting Abstracts \#50},
  volume~50 of \emph{AAS/Division for Planetary Sciences Meeting Abstracts},
  page 408.08, October 2018.

\bibitem[{Moyer}(2000)]{Moyer2000}
T.D. {Moyer}.
\newblock Formulation for observed and computed values of deep space network
  data types for navigation.
\newblock Monography of DEEP SPACE COMMUNICATIONS AND NAVIGATION Series~2, JPL,
  2000.

\bibitem[{Park} et~al.(2017){Park}, {Folkner}, {Konopliv}, {Williams}, {Smith},
  and {Zuber}]{2017AJ....153..121P}
R.~S. {Park}, W.~M. {Folkner}, A.~S. {Konopliv}, J.~G. {Williams}, D.~E.
  {Smith}, and M.~T. {Zuber}.
\newblock {Precession of Mercury's Perihelion from Ranging to the MESSENGER
  Spacecraft}.
\newblock \emph{\aj}, 153:\penalty0 121, March 2017.
\newblock \doi{10.3847/1538-3881/aa5be2}.

\bibitem[{Pijpers}(1998)]{1998MNRAS.297L..76P}
F.~P. {Pijpers}.
\newblock {Helioseismic determination of the solar gravitational quadrupole
  moment}.
\newblock \emph{\mnras}, 297\penalty0 (3):\penalty0 L76--L80, July 1998.
\newblock \doi{10.1046/j.1365-8711.1998.01801.x}.

\bibitem[{Roca Cort{\'e}s} et~al.(1998){Roca Cort{\'e}s}, {Lazrek}, {Bertello},
  {Thiery}, {Baudin}, {Garcia}, and {GOLF Team}]{1998ESASP.418..329R}
T.~{Roca Cort{\'e}s}, M.~{Lazrek}, L.~{Bertello}, S.~{Thiery}, F.~{Baudin},
  R.~A. {Garcia}, and {GOLF Team}.
\newblock {The Solar Acoustic Spectrum as Seen by GOLF. III. Asymmetries,
  Resonant Frequencies and Splittings}.
\newblock In S.~{Korzennik}, editor, \emph{Structure and Dynamics of the
  Interior of the Sun and Sun-like Stars}, volume 418 of \emph{ESA Special
  Publication}, page 329, January 1998.

\bibitem[{Roxburgh}(2001)]{2001A&A...377..688R}
I.~W. {Roxburgh}.
\newblock {Gravitational multipole moments of the Sun determined from
  helioseismic estimates of the internal structure and rotation}.
\newblock \emph{\aap}, 377:\penalty0 688--690, October 2001.
\newblock \doi{10.1051/0004-6361:20011104}.

\bibitem[{Scherrer} and {Gough}(2019)]{2019ApJ...877...42S}
P.~H. {Scherrer} and D.~O. {Gough}.
\newblock {A Critical Evaluation of Recent Claims Concerning Solar Rotation}.
\newblock \emph{\apj}, 877\penalty0 (1):\penalty0 42, May 2019.
\newblock \doi{10.3847/1538-4357/ab13ad}.

\bibitem[{Schunker} et~al.(2018){Schunker}, {Schou}, {Gaulme}, and
  {Gizon}]{2018SoPh..293...95S}
Hannah {Schunker}, Jesper {Schou}, Patrick {Gaulme}, and Laurent {Gizon}.
\newblock {Fragile Detection of Solar g-Modes by Fossat et al.}
\newblock \emph{\solphys}, 293\penalty0 (6):\penalty0 95, June 2018.
\newblock \doi{10.1007/s11207-018-1313-6}.

\bibitem[{Soffel} and {Frutos}(2016)]{2016JGeod..90.1345S}
M.~{Soffel} and F.~{Frutos}.
\newblock {On the usefulness of relativistic space-times for the description of
  the Earth's gravitational field}.
\newblock \emph{Journal of Geodesy}, 90\penalty0 (12):\penalty0 1345--1357,
  December 2016.
\newblock \doi{10.1007/s00190-016-0927-4}.

\bibitem[{Standish}(2001)]{Standish2001}
E.M. {Standish}.
\newblock The jpl de405 planetary and lunar ephemerides.
\newblock 2001.

\bibitem[{Thompson} et~al.(2003){Thompson}, {Christensen-Dalsgaard}, {Miesch},
  and {Toomre}]{2003ARA&A..41..599T}
M.~J. {Thompson}, J.~{Christensen-Dalsgaard}, M.~S. {Miesch}, and J.~{Toomre}.
\newblock {The Internal Rotation of the Sun}.
\newblock \emph{\araa}, 41:\penalty0 599--643, January 2003.
\newblock \doi{10.1146/annurev.astro.41.011802.094848}.

\bibitem[{Thor} et~al.(2020){Thor}, {Kallenbach}, {Christensen}, {Stark},
  {Steinbr\"ugge}, {Di Ruscio}, {Cappuccio}, {Iess}, {Hussmann}, and
  {Oberst}]{thor20}
R.~N. {Thor}, R.~{Kallenbach}, U.~R. {Christensen}, A.~{Stark},
  G.~{Steinbr\"ugge}, A.~{Di Ruscio}, P.~{Cappuccio}, L.~{Iess}, H.~{Hussmann},
  and J.~{Oberst}.
\newblock Prospects for measuring mercury\'{}s tidal love number h2 with the
  bepicolombo laser altimeter.
\newblock \emph{A\&A}, 633:\penalty0 A85, 2020.
\newblock \doi{10.1051/0004-6361/201936517}.
\newblock URL \url{https://doi.org/10.1051/0004-6361/201936517}.

\end{thebibliography}
\end{document}